\documentclass[a4paper,twocolumn,10pt,accepted=2022-08-18]{quantumarticle}
\pdfoutput=1
\usepackage[utf8]{inputenc}
\usepackage[english]{babel}
\usepackage[T1]{fontenc}
\usepackage{amsmath}
\usepackage{amssymb}
\usepackage{hyperref}
\usepackage{cite}
\usepackage{url}

\begin{document}

\title{Superfluid drag between excitonic polaritons and superconducting electron gas}

\author{Azat F. Aminov}
\email{afaminov@hse.ru}
\affiliation{National Research University Higher School of Economics, 109028 Moscow, Russia}

\author{Alexey A. Sokolik}
\email{asokolik@hse.ru}
\affiliation{Institute for Spectroscopy, Russian Academy of Sciences, 142190 Troitsk, Moscow, Russia}
\affiliation{National Research University Higher School of Economics, 109028 Moscow, Russia}

\author{Yurii E. Lozovik}
\email{lozovik@isan.troitsk.ru}
\affiliation{Institute for Spectroscopy, Russian Academy of Sciences, 142190 Troitsk, Moscow, Russia}
\affiliation{National Research University Higher School of Economics, 109028 Moscow, Russia}

\maketitle

\begin{abstract}
The Andreev-Bashkin effect, or superfluid drag, is predicted in a system of Bose-condensed excitonic polaritons in optical microcavity coupled by electron-exciton interaction with a superconducting layer. Two possible setups with spatially indirect dipole excitons or direct excitons are considered. The drag density characterizing a magnitude of this effect is found by many-body calculations with taking into account dynamical screening of electron-exciton interaction. For the superconducting electronic layer, we assume the recently proposed polaritonic mechanism of Cooper pairing, although the preexisting thin-film superconductor should also demonstrate the effect. According to our calculations, the drag density can reach considerable values in realistic conditions, with excitonic and electronic layers made from GaAs-based quantum wells or two-dimensional transition metal dichalcogenides. The predicted nondissipative drag could be strong enough to be observable as induction of a supercurrent in the electronic layer by a flow of polariton Bose condensate.
\end{abstract}

\section{Introduction}

Drag effects consisting in interaction-induced mutual entrainment between different kinds of particles in two-component or bilayer systems were extensively studied in semiconductor heterostructures \cite{Rojo1999,Narozhny2016}. Recent observations of the anomalous enhancement of the drag in electron-hole systems \cite{Morath2009,Croxall2008,Li2017}, which is probably caused by electron-hole Cooper pairing or exciton superfluidity \cite{Lozovik1976,Efimkin2016}, poses new challenges to this field and present a spectacular example of how quantum coherent effects manifest themselves in transport phenomena. Similar anomalies were observed in the drag between normal metallic and superconducting films \cite{Giordano1994,Huang1995,JC2020}.

When both components are superconducting or superfluid, a  nondissipative drag effect is possible, when a non-zero flow of one superfluid component induces a non-zero supercurrent of the other one. Such superfluid drag, or Andreev-Bashkin effect (ABE),  was initially predicted for a mixture of superfluid ${}^3\mathrm{He}$ and ${}^4\mathrm{He}$ \cite{AB}, although it was never observed in this system due to low miscibility of the constituents. The similar nondissipative drag effect was predicted for superconductors \cite{Duan1993,Rojo1999}. Nevertheless, there is no yet direct and unambiguous observation of the superfluid drag effect (see \cite{Khalid2021} and references therein).

Theoretical proposals predict signatures of ABE in cold atomic gas mixtures in a shift of dipole oscillation frequency in a trap, as studied analytically \cite{Fil2005, Romito2020, MikiOta2020, Tanatar2020} and by means of Monte-Carlo simulations \cite{Selin2018,Stian2018,Nespolo2018}, or in appearance of a phase shift across barrier, as could be detected via atomic interferometry \cite{Khalid2021}. ABE can also evince itself in formation and behavior of quantum vertices, modifying the Berezinskii-Kosterlitz-Thouless transition in two-dimensional mixtures of gases \cite{Karle2019} and playing presumably important role in the dynamics of superfluid mixtures in neutron stars \cite{Alpar1984,Babaev2004}. It should be noted that in spatially nonuniform systems mean-field interaction effects in a two-component Bose condensate (or between two neighboring condensates) can induce another kind of superfluid drag between the components competing with ABE \cite{Lozovik2002}, so its careful discrimination from ABE should be performed in experiments \cite{Khalid2021}.

Modern two-dimensional materials and heterostructures \cite{Novoselov2016,Vincent2021} provide a prospective platform for realization of drag effects. Hybrid Bose-Fermi systems with polaronic and drag effects induced by electron-exciton and electron-polariton interactions are especially interesting in this context \cite{Lozovik1997,Lozovik1999,Boev2019,Berman2010,Cotlet(2019)}. Formation of polaritons in semiconductor quantum wells and two-dimensional transition metal dichalcogenides embedded into optical microcavities \cite{QFl} allows to enhance tunability of the Bose-Fermi systems, to increase the critical temperature of Bose-Einstein condensation (BEC) and to employ polaronic effects. For instance, drag between Bose-condensed polaritons and electrons in normal state, both located in the same layer, was observed \cite{Myers2021} and theoretically explained \cite{Mukherjee2022}.

Recently the novel mechanism of superconductivity has been proposed \cite{Laussy2010, Laussy2012, Cotlet2016, Petros2018, Cherotchenko2016, Sedov2019, Meng2021, Sun2021_1, Sun2021}, when electrons in a two-dimensional electron gas (2DEG) undergo Cooper pairing due to exchange of virtual Bogoliubov excitations in BEC of excitonic polaritons. The resulting superfluid Bose-Fermi system could be a natural platform to observe signatures of ABE between superfluid polariton BEC and superconducting 2DEG. The structure of similar geometry with a GaAs-based 2DEG layer embedded into a microcavity was realized in the recent experiment \cite{Sven2021}.

In this paper we consider ABE between BEC of excitonic polaritons in optical microcavity and 2DEG in a superconducting state, separated by a distance of the order of 10-100 nm from the excitonic layer. For the latter, we consider two different setups with spatially indirect and direct excitons, as shown in Fig.~\ref{fig:ExpSet}. ABE manifests itself as appearance of a superconducting current in the electronic layer with the mass current density $\mathbf{g}_\mathrm{el}=\rho_\mathrm{dr}\mathbf{v}_\mathrm{p}$ induced by the nonzero velocity $\mathbf{v}_\mathrm{p}$ of the polariton BEC. Alternatively, a nonzero velocity $\mathbf{v}_\mathrm{el}$ of the Cooper pair condensate in the electronic layer can induce the flow of polariton superfluid component with the mass current density $\mathbf{g}_\mathrm{p}=\rho_\mathrm{dr}\mathbf{v}_\mathrm{el}$ \cite{AB}. The coefficient $\rho_\mathrm{dr}$, called the superfluid drag density, determines a magnitude of ABE.

We assume that superconductivity in the electronic layer is induced by the polaritonic-BEC mechanism proposed in Refs. \cite{Laussy2010, Laussy2012, Cherotchenko2016, Cotlet2016, Petros2018, Sedov2019, Meng2021, Sun2021_1, Sun2021}, although the specific pairing mechanism is not important for our analysis. In Sec.~\ref{Sec2} we describe our theoretical approach for calculating the superfluid drag density $\rho_\mathrm{dr}$ using the many-body theory. BEC of polaritons and superconducting 2DEG are described using, respectively, Bogoliubov and Bardeen-Cooper-Schrieffer (BCS) approaches. The important ingredient of the theory is taking into account the screening of the interlayer electron-exciton interaction causing ABE.

In Sec. \ref{Sec3} we present the results of our numerical calculations of the drag density $\rho_\mathrm{dr}$ in realistic conditions, where both excitonic and electronic layers can be based on GaAs/AlGaAs quantum wells (QW) or two-dimensional semiconducting transition metal dichalcogenides (TMDC). Our results show that  $\rho_\mathrm{dr}$ can reach 0.001-0.1 of the total mass density of polaritons, so ABE can be strong enough to be observable. We propose the observation method when the polariton flow is created and induced nondissipative current in the superconducting layer is detected. In Sec. \ref{Sec4} we state our conclusions.

\section{Theory}\label{Sec2}
\subsection{System overview}

\begin{figure}[t]
\centering
\includegraphics[width=\columnwidth]{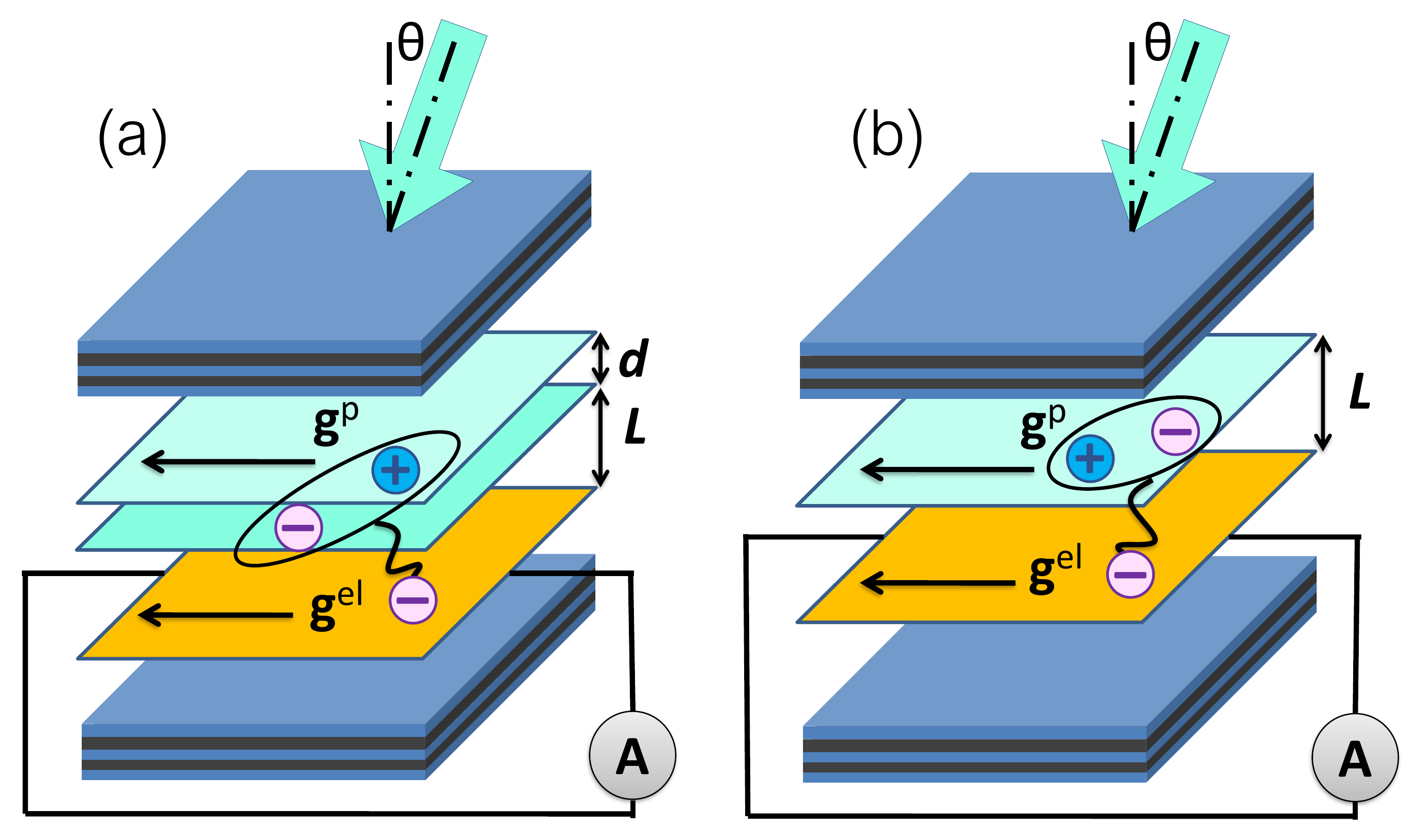}
\caption{System schematic: two-layer polariton-2DEG system with indirect (a) and direct (b) excitons forming polaritons by hybridizing with microcavity photons. Electron-exciton interaction couples superfluid polariton and superconducting electron condensates, giving rise to the Andreev-Bashkin effect. Creating a flow $\mathbf{g}^{\mathrm{p}}$ of polariton superfluid component by, for example, inclined incident laser beam and measuring the induced superconducting current with the mass density $\mathbf{g}^\mathrm{el}$ in 2DEG would allow to observe the effect.}
\label{fig:ExpSet}
\end{figure}

Spatially indirect polariton (or dipolariton) has a constant electrical dipole moment, which rises from an exciton. This dipole moment is caused either by spatial separation of electrons and holes in a single QW using a perpendicular electric field or via doping two coupled QWs with electrons and holes \cite{Snoke2010}. Without loss of generality, we consider the latter case, as shown in Fig. \ref{fig:ExpSet}(a). Experimentally indirect excitons were observed in TMDC bilayer structures \cite{Calman2018,Datta2021}, coupled QWs \cite{Butov1999,dipdip2} and wide single QWs \cite{Miller1985,Polland1985}.

On the other hand, direct polariton is a superposition of a photon and direct exciton, which is a bound state of electron and hole located in the same QW \cite{Kasprzak2006} or TMDC layer \cite{Zhao2021} and thus do not possess a constant dipole moment (Fig. \ref{fig:ExpSet}(b)). Although the dipole-electron interaction between indirect polaritons and electrons is expected to be stronger, the coupling of indirect excitons with light is weaker than that of direct excitons, and so far a dipolariton BEC has not been observed \cite{dipdip,dipdip2}. In contrast, BEC of direct polaritons is routinely obtained in experiments \cite{QFl,Zhao2021,Kasprzak2006}, although the interaction of direct excitons with electrons is much weaker so the drag effects in such systems could be less pronounced.

According to the theory and experiment \cite{dipdip, dipdip2}, the microcavity systems with indirect polaritons demonstrate coexistence of both direct and indirect excitons, coupled to light and to each other. However, similarly to Refs. \cite{Laussy2010, Laussy2012, Cotlet2016, Meng2021}, we do not account for such details and assume a simplified two-oscillator model of indirect excitons coupled with the cavity photons.

The conventional definition of superfluidity as the absence of dissipative friction is questionable for nonequilibrium systems of polaritons with a finite lifetime \cite{Wouters2010,Szymaska2006}. For instance, currents in Bose-condensed polariton systems always undergo drag force on impurities even at velocities lower than the Landau critical velocity \cite{Wouters2010}, although this force substantially decreases in the presence of BEC. Despite that, exciton-polariton systems in experiments exhibit multiple characteristic properties of superfluids: suppression of scattering on obstacles \cite{Lerario2017,Amo2009}, dissipationless long-range propagation \cite{Snoke2013}, and quasi long-range order in both spatial and temporal coherences \cite{Caputo2017}. Anyway, our setup does not require robustness of the polaritonic superfluidity, it needs only emergence of polariton BEC and the approximate persistence of its condensate velocity $\mathbf{v}_\mathrm{p}$ in the region of electron-polariton interaction. Therefore, throughout the paper we consider the ``superfluid'' motion of polaritons participating in the superfluid drag as the motion of their Bose condensate which can be considered as homogeneous on spatial and time scales larger than those of internal electron-polariton system dynamics. Additionally, we do not consider an influence of disorder-induced mesoscopic condensate inhomogeneities, vortices and finite-size effects on ABE, since these topics are beyond the scope of our paper.

The Hamiltonian of two-layered exciton-electron system, embedded in microcavity, reads
\begin{multline}\label{H}
H=\sum_{\mathbf{k}} \Psi^{\dag}_{\mathbf{k}}H_{0}\Psi_{\mathbf{k}}+\sum_{\mathbf{k}s} \epsilon^{\mathrm{el}}_{\mathbf{k}} a^{\dag}_{\mathbf{k}s}a_{\mathbf{k}s}\\ + H_{\mathrm{int}}^{\mathrm{el-el}}+H_{\mathrm{int}}^{\mathrm{el-x}}+H_{\mathrm{int}}^{\mathrm{x-x}},
\end{multline}
where $\Psi_{\mathbf{k}}=(\gamma_{\mathbf{k}},c_{\mathbf{k}})^{T}$ is the column of destruction operators of photon $(\gamma_{\mathbf{k}})$ and exciton ($c_{\mathbf{k}}$), $a_{\mathbf{k}s}$ is the destruction operator of electron with momentum $\mathbf{k}$ and spin $s$. The matrix $H_{0}$ is the exciton-photon Hamiltonian without the exciton-exciton interaction:
\begin{equation}
H_{0} = \left(\begin{array}{cc}
\epsilon^{\mathrm{c}}_{\mathbf{k}} & \Omega_{\mathrm{R}} \\
\Omega_{\mathrm{R}} & \epsilon^{\mathrm{x}}_\mathbf{k}
\end{array}\right),
\end{equation}
where $\epsilon^{\mathrm{c}}_{\mathbf{k}} = k^{2}/2m_{\mathrm{c}}+\delta$, $\epsilon^{\mathrm{x}}_{\mathbf{k}} = k^{2}/2m_{\mathrm{x}}$ are the bare dispersions of photons and excitons, $m_{\mathrm{c}}$ and $m_{\mathrm{x}}$ are their effective masses, $\delta$ is the photon-to-exciton detuning, $\Omega_{\mathrm{R}}$ is the Rabi splitting. Hereafter we put $\hbar = 1$. The term $H_\mathrm{int}^\mathrm{el-el}$ is the standard Coulomb interaction of electrons, and the other interaction terms in Eq. (\ref{H}) are:
\begin{align}\label{Hint}
H_{\mathrm{int}}^{\mathrm{x-x}} = \frac{1}{2S}\sum_{\mathbf{kk^{\prime}q}} g^{\mathrm{x-x}} c^{\dag}_{\mathbf{k+q}}c^{\dag}_{\mathbf{k^{\prime}-q}}c_{\mathbf{k^{\prime}}}c_{\mathbf{k}}, \\ \label{Hint_el_x}
H_{\mathrm{int}}^{\mathrm{el-x}} = \frac{1}{S}\sum_{\mathbf{kk^{\prime}q}s} V^{\mathrm{el-x}}(q) c^{\dag}_{\mathbf{k+q}}a^{\dag}_{\mathbf{k^{\prime}-q},s}a_{\mathbf{k^{\prime}}s}c_{\mathbf{k}},
\end{align}
where $g^{\mathrm{x-x}}$ is a constant of interaction between excitons, which, similarly to \cite{Laussy2010, Laussy2012, Cherotchenko2016, Cotlet2016, Petros2018, Meng2021}, is assumed to be independent of momentum; $V^{\mathrm{el-x}}$ is the interaction between electrons and excitons, and $S$ is the quantization area. Next, we conventionally perform diagonalization of the first term in Eq.~(\ref{H}), introducing new quasiparticles, namely lower ($b_{\mathbf{k}}$) and upper ($\tilde{b}_{\mathbf{k}}$) polaritons:
\begin{equation}
\left(\begin{array}{cc}
b_{\mathbf{k}} \\
\tilde{b}_{\mathbf{k}}
\end{array}\right) =
\left(\begin{array}{cc}
\sqrt{1-X_{k}^{2}} & X_{k} \\
-X_{k} & \sqrt{1-X_{k}^{2}}
\end{array}\right)
\left(\begin{array}{cc}
\gamma_{\mathbf{k}} \\
c_{\mathbf{k}}
\end{array}\right),
\end{equation}
with the Hopfield coefficient $X_{k}$. BEC in equilibrium occurs only on the lower polariton branch, so, similarly to \cite{Laussy2010, Laussy2012, Cotlet2016, Petros2018, Meng2021}, we truncate the Hamiltonian to the lower polaritons with the dispersion
\begin{multline}\label{Epol}
\epsilon^{\mathrm{p}}_{\mathbf{k}} =\frac{1}{2}\left[\sqrt{\delta^{2}+4 \Omega_{\mathrm{R}}^{2}}+\frac{ k^{2}}{2 m_{\mathrm{p}}} \right. \\ \left. - \sqrt{(\epsilon^{\mathrm{c}}_{\mathbf{k}} - \epsilon^{\mathrm{x}}_{\mathbf{k}})^{2}+4 \Omega_{\mathrm{R}}^{2}}\right],
\end{multline}
where the inverse polariton mass is $m_{\mathrm{p}}^{-1}=m_{\mathrm{x}}^{-1}+m_{\mathrm{c}}^{-1}$. The interaction Hamiltonians (\ref{Hint})--(\ref{Hint_el_x}) in terms of the lower polariton operators read
\begin{align}
H_{\mathrm{int}}^{\mathrm{x-x}} = \frac{1}{2S}\sum_{\mathbf{kk^{\prime}q}}g^{\mathrm{p-p}}_{\mathbf{kk}^{\prime}}(\mathbf{q}) b^{\dag}_{\mathbf{k+q}}b^{\dag}_{\mathbf{k^{\prime}-q}}b_{\mathbf{k^{\prime}}}b_{\mathbf{k}}, \\
H_{\mathrm{int}}^{\mathrm{el-x}} = \frac{1}{S}\sum_{\mathbf{kk^{\prime}q}s} V^{\mathrm{el-p}}_\mathbf{k}(\mathbf{q}) b^{\dag}_{\mathbf{k+q}}a^{\dag}_{\mathbf{k^{\prime}-q},s}a_{\mathbf{k^{\prime}}s}b_{\mathbf{k}}.
\end{align}
We introduced the polariton-polariton interaction
\begin{equation}\label{gppBorn}
g^{\mathrm{p-p}}_{\mathbf{kk}^{\prime}}(\mathbf{q}) = X_{k}X_{k^{\prime}}X_{|\mathbf{k}^{\prime}-\mathbf{q}|}X_{|\mathbf{k}+\mathbf{q}|} g^{\mathrm{x-x}},
\end{equation}
and the unscreened electron-polariton interaction
\begin{equation}\label{VepBorn}
V^{\mathrm{el-p}}_\mathbf{k}(\mathbf{q})=X_{k}X_{|\mathbf{k}+\mathbf{q}|}V^{\mathrm{el-x}}(q).
\end{equation}
Such dressing of interactions with the Hopfield coefficients corresponds to the Born approximation. Recent theoretical studies \cite{HuiHu2020, Bleu2020, Li2021} has shown that, in the presence of the strong light-matter coupling, interactions involving 2D polaritons needs to be treated beyond this approximation, and actual magnitudes of interactions turns out to be lower than given by Eqs. (\ref{gppBorn})--(\ref{VepBorn}). However, we still use these conventional expressions, since they are widely accepted and adequately describe most of the experimental results (see \cite{Estrecho2019} and references therein).

Regarding to the BEC of polaritons, we assume that its mean-field critical temperature is much higher than the system temperature, $T^{\mathrm{BEC}}_{\mathrm{c}}\gg T$, and the density of non-condensate particles is negligible: $n^{\mathrm{p}} \approx n_{0}^{\mathrm{p}}$; here $n^{\mathrm{p}}$ is the total polariton density and $n_{0}^{\mathrm{p}}$ is the condensate density. This assumption allows us to use the Bogoliubov theory, which is conventionally applied to describe BEC of excitonic polaritons in a microcavity \cite{Utsunomia2008}.

The electron layer is assumed to be a 2D superconductor with the s-wave pairing and the zero-temperature energy gap ${\Delta \approx 1.76 T^{\mathrm{SC}}_{\mathrm{c}}}$, where ${T^{\mathrm{SC}}_{\mathrm{c}}}$ is the superconducting transition temperature. We assume that the superconductivity is induced by exchange of Bogoliubov excitations in the polariton BEC, as proposed in Refs. \cite{Laussy2010, Laussy2012, Cherotchenko2016, Cotlet2016, Petros2018, Sedov2019, Meng2021, Sun2021_1, Sun2021}. However, the specific mechanism of electron pairing is not essential for our analysis: superconducting phase can be result of interaction between electrons and either polaritons or, for instance, phonons, excitons or spin fluctuations. We just suppose that $T^{\mathrm{SC}}_{\mathrm{c}}$ is much lower than the Fermi energy of electron system and use the weak-coupling BCS theory to describe the superconducting 2DEG.

In the theory of Coulomb drag between layers of normal metals \cite{Narozhny2016}, accounting for the sources of dissipation such as interactions with impurities or phonons is essential. Two major regimes are usually distinguished: the ballistic one, when the mean free path of particles is longer than the interlayer distance, and the diffusive one in the opposite case. Contrary to the normal Coulomb drag, dissipation processes are not crucial for existence of the superfluid drag effect \cite{Fil2005,Romito2020}. Moreover, recent progress in material science allowed to create very clean materials, in which the mean free path of both electrons and polaritons is large enough to justify assumption of the the ballistic regime in calculations. For instance, in this paper we assume the interlayer distance to be about $10\,\mbox{nm}$, while the mean free path of electrons in typical TMDC, MoS$_{2}$, can exceed 20 nm \cite{MoS2}.

\begin{figure}[t]
\centering
\includegraphics[width=\columnwidth]{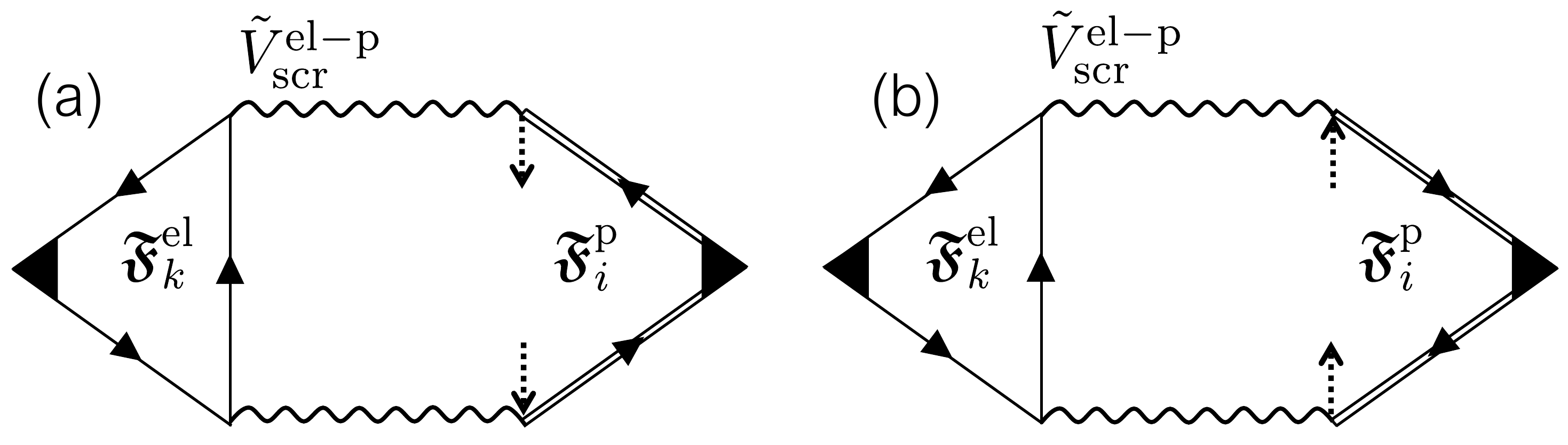}
\caption{The main diagrams contributing to the interlayer current-current response function (\ref{Correlator}) and hence to the drag density $\rho_\mathrm{dr}$. Single and double straight lines denote electron and polariton Green functions, wavy lines is the interlayer electron-polariton interaction, and filled triangles at the ends are the current vertices $\mathbf{g}^{l}$. Dotted arrows denote the condensate of polaritons $\sqrt{n_{0}^{\mathrm{p}}}$.}
\label{fig: Diagram for EP}
\end{figure}

\subsection{Calculation of drag density}
Our main goal is to calculate the superfluid drag density. In the linear response theory, it can be found from the correlation function of currents \cite{Romito2020}:
\begin{equation}\label{drag}
\rho_{\mathrm{dr}} = - \lim_{k \to 0}  \chi^{\mathrm{el-p}}_{\mathrm{T}}(k,0).
\end{equation}
Here $\chi_{\mathrm{T}}$ is a transverse part of the interlayer current-current response tensor
\begin{multline}\label{Correlator}
\chi_{ik}^{\mathrm{el-p}}(\mathbf{k}, i\omega) =\frac1S\int_{-1/T}^{1/T} d \tau e^{i \omega\left(\tau-\tau^{\prime}\right)} \\
\times\left\langle T_{\tau} \mathbf{g}^{\mathrm{el}}_{i}(\mathbf{k}, \tau) \mathbf{g}^{\mathrm{p}}_{k}\left(-\mathbf{k},\tau^{\prime}\right)\right\rangle,
\end{multline}
where $i,k = x,y$. The Fourier harmonics of the mass current densities $\mathbf{g}^\mathrm{el}$ and $\mathbf{g}^\mathrm{p}$ in, respectively, electron and polariton subsystems, are given by the operators $\mathbf{g}^{\mathrm{el}}(\mathbf{k})=g_\mathrm{v}\sum_{\mathbf{p}s} (\mathbf{p+k}/2) a^{\dag}_{\mathbf{p}s}a_{\mathbf{p+k},s}$, ${\mathbf{g}^{\mathrm{p}}(\mathbf{k})=\sum_{\mathbf{p}} (\mathbf{p+k}/2) b^{\dag}_{\mathbf{p}}b_{\mathbf{p+k}}}$, where $g_\mathrm{v}$ is the valley degeneracy factor (1 for QW and 2 for TMDC electronic layer). Following a standard calculation based on the Matsubara technique \cite{Narozhny2016}, it can be shown that in a clean system or with uncorrelated disorder potentials in both layers the first order in the expansion of $\chi^{\mathrm{el-p}}_{ik}$ over the interlayer electron-polariton interaction is zero in the static limit $i\omega\to0$, and the leading order is given by the second-order term \cite{Boev2019}:
\begin{multline}\label{chidiag}
\chi^{\mathrm{el-p}}_{ik}(\mathbf{0},0)=\frac{T}{2 S}\sum_{\mathbf{q},i\omega_{m}} \boldsymbol{\mathfrak{F}}_{i}^{\mathrm{p}}(\mathbf{q},i\omega_{m}) \boldsymbol{\mathfrak{F}}_{k}^{\mathrm{el}}(\mathbf{q},i\omega_{m}) \\ \times \left|\tilde{V}^{\mathrm{el-p}}_{\mathrm{scr}}(q,i\omega_{m})\right|^{2}.
\end{multline}
Here $\boldsymbol{\mathfrak{F}}_{k}^{\mathrm{el}}$ and $\boldsymbol{\mathfrak{F}}_{i}^{\mathrm{p}}$ are the nonlinear current-density response, or rectification, functions of, respectively, superconducting 2DEG and Bose-condensed polariton system, $\tilde{V}^{\mathrm{el-p}}_{\mathrm{scr}}(q,i\omega_{m})$ is the partially screened electron-polariton interaction (see Eq.~(\ref{TildeVel-p}) below); $i\omega_{m}=2\pi imT$ are bosonic Matsubara frequencies. The corresponding Feynman diagrams accounting for two contributions to $\boldsymbol{\mathfrak{F}}_{k}^{\mathrm{el}}$ and $\boldsymbol{\mathfrak{F}}_{i}^{\mathrm{p}}$ are depicted in Fig.~\ref{fig: Diagram for EP}. We do not take into account the conventional diagrams of the normal drag theory \cite{Narozhny2016} with triangles of noncondensate polariton Green functions, since, according to our numerical estimates given in Appendix~\ref{Appendix_A}, their contribution at $T \ll T^{\mathrm{BEC}}_{\mathrm{c}}$ is negligible.

\subsection{Nonlinear response functions}
\subsubsection{Polaritons}\label{secPol}

To calculate the rectification function $\boldsymbol{\mathfrak{F}}^{\mathrm{p}}$ of polaritons, we use the conventional normal and anomalous Green functions of Bose-condensed gas in the Bogoliubov approach \cite{Landau} with accounting for the dependence of polariton-polariton interaction $g^{\mathrm{p-p}}_{\mathbf{kk}^{\prime}}(\mathbf{q})$ (\ref{gppBorn}) on three momenta. The mean-field contributions to the normal and anomalous self-energies are, respectively, $\Sigma^{\mathrm{n}}(q) = n_{0}^{\mathrm{p}}[g^{\mathrm{p-p}}_{0\mathbf{q}}(\mathbf{q})+g^{\mathrm{p-p}}_{0\mathbf{q}}(0)]$ and $\Sigma^{\mathrm{a}}(q) = n_{0}^{\mathrm{p}}g^{\mathrm{p-p}}_{0\mathbf{q}}(\mathbf{q})$ \cite{Landau}. The matrix Green function of polaritons $\hat{G}^{\mathrm{p}}(\mathbf{q},\tau) = - \langle T_{\tau} B(\mathbf{q},\tau)B^{\dag}(\mathbf{q},0) \rangle$, where $B(\mathbf{q},\tau) = (b_\mathbf{q}(\tau),b^{\dag}_{-\mathbf{q}}(\tau) )^{\mathrm{T}}$, is found from the Dyson-Beliaev equations:
\begin{multline}\label{GreenPolariton}
\hat{G}^{\mathrm{p}} (\mathbf{q},i\omega) = \frac{1}{(i\omega)^{2} - \left(E^{\mathrm{p}}_{\mathbf{q}}\right)^{2}} \\ \times \left(\begin{array}{cc}
i\omega + \tilde{\epsilon}^{\mathrm{p}}_{\mathbf{q}} + c_{q} & -c_{q} \\
-c_{q} & -i\omega + \tilde{\epsilon}^{\mathrm{p}}_{\mathbf{q}} + c_{q}
\end{array}\right),
\end{multline}
where $\tilde{\epsilon}^{\mathrm{p}}_{\mathbf{q}} = \epsilon^{\mathrm{p}}_{\mathbf{q}} + c_{q} - c_{0} $. Here we have subtracted the chemical potential $c_{0} = X_{0}^{4}n_{0}^{\mathrm{p}}g^{\mathrm{x-x}}$ to make the Bogoliubov quasi-particle dispersion $E^{\mathrm{p}}_{\mathbf{q}} = \sqrt{\tilde{\epsilon}^{\mathrm{p}}_{\mathbf{q}}\left(\tilde{\epsilon}^{\mathrm{p}}_{\mathbf{q}}+2c_{q}\right)}$ gapless, as required by the Goldstone theorem; $c_{q} =X_{q}^{2} X_{0}^{2}n_{0}^{\mathrm{p}}g^{\mathrm{x-x}}$ is the Hartree part of the normal self-energy $\Sigma^\mathrm{n}(q)$.

With the Green functions (\ref{GreenPolariton}), similarly to \cite{Boev2019}, we write the lowest order contribution of the processes involving the condensate to the rectification function. It is proportional to the density of polariton condensate $n_0^\mathrm{p}$:
\begin{multline}\label{Fdia}
\boldsymbol{\mathfrak{F}}^{\mathrm{p}}(\mathbf{q},i\omega_{m})= n_{0}^{\mathrm{p}} \mathbf{q} \left\{\hat{G}_{11}^{\mathrm{p}}(\mathbf{q},i\omega_{m})+\hat{G}_{12}^{\mathrm{p}} (\mathbf{q},i\omega_{m})\right\}^{2}\\ + (\mathbf{q},i\omega_{m}\to-\mathbf{q},-i\omega_{m}).
\end{multline}
The diagrams corresponding to this expression are shown in the right parts of Figs.~\ref{fig: Diagram for EP}(a,b).

Substituting (\ref{GreenPolariton}) into (\ref{Fdia}), we find that the rectification function of the polariton system within the Bogoliubov theory has the same form as the exciton one \cite{Boev2019} and reads
\begin{equation}\label{Fb}
\boldsymbol{\mathfrak{F}}^{\mathrm{p}}(\mathbf{q},i\omega_{m}) = \frac{4\mathbf{q} \, i\omega_{m} \, n_0^\mathrm{p}\tilde\epsilon^\mathrm{p}_\mathbf{q}}{\left[(i\omega_{m})^{2}-\left(E^{\mathrm{p}}_{\mathbf{q}} \right)^{2} \right]^{2}}.
\end{equation}
As shown in Appendix~\ref{Appendix_A}, the noncondensate contributions to $\boldsymbol{\mathfrak{F}}^{\mathrm{p}}$, given by the triangular diagrams with three polaritonic Green functions, are negligible at the assumed low temperature.

\subsubsection{Superconductor}\label{secSuper}
To calculate the rectification function of the superconducting 2DEG $\boldsymbol{\mathfrak{F}}^{\mathrm{el}}$, we use the Nambu formalism of the matrix Green function \cite{Nambu1960,Schrieffer1970} of superconductor: $\hat{G}^{\mathrm{el}}(\mathbf{q},\tau) = - \langle T_{\tau} A(\mathbf{q},\tau)A^{\dag}(\mathbf{q},0) \rangle$, where $A(\mathbf{q},\tau) = (a_{\mathbf{q}\uparrow}(\tau),a^{\dag}_{-\mathbf{q}\downarrow}(\tau) )^{\mathrm{T}}$. In the BCS theory, the Green functions are \cite{Schrieffer1970}:
\begin{multline}\label{GreenBCS}
\hat{G}^{\mathrm{el}} (\mathbf{q},i\omega) \\ =  \left(\begin{array}{cc}
\frac{u_{\mathbf{p}}^{2}}{i\omega-E_{\mathbf{p}}}+\frac{v_{\mathbf{p}}^{2}}{i\omega+E_{\mathbf{p}}} & \frac{u_{\mathbf{p}} v_{\mathbf{p}}}{i\omega-E_{\mathbf{p}}}-\frac{u_{\mathbf{p}} v_{\mathbf{p}}}{i\omega+E_{\mathbf{p}}} \\
\frac{u_{\mathbf{p}} v_{\mathbf{p}}}{i\omega-E_{\mathbf{p}}}-\frac{u_{\mathbf{p}} v_{\mathbf{p}}}{i\omega+E_{\mathbf{p}}} & \frac{v_{\mathbf{p}}^{2}}{i\omega-E_{\mathbf{p}}}+\frac{u_{\mathbf{p}}^{2}}{i\omega+E_{\mathbf{p}}}
\end{array}\right).
\end{multline}
Here $u_{\mathbf{p}}^{2},v_{\mathbf{p}}^{2}=\frac{1}{2}\left(1\pm\epsilon^{\mathrm{el}}_{\mathbf{p}} / E_{\mathbf{p}}^{\mathrm{el}}\right)$, $u_{\mathbf{p}} v_{\mathbf{p}}=\Delta / 2 E_{\mathbf{p}}^{\mathrm{el}}, E_{\mathbf{p}}^{\mathrm{el}}=\sqrt{\left(\epsilon^{\mathrm{el}}_{\mathbf{p}}\right)^{2}+\Delta^{2}}$, $\epsilon^{\mathrm{el}}_{\mathbf{p}} = p^{2}/2m^{*}_{\mathrm{e}} - E_{\mathrm{F}}$ is the bare electron dispersion, $m^{*}_{\mathrm{e}}$ and $E_{\mathrm{F}}$ are the effective mass of electrons and the Fermi energy of 2DEG.

The diagrams for $\boldsymbol{\mathfrak{F}}^{\mathrm{el}}$, shown in the left parts of Figs.~\ref{fig: Diagram for EP}(a,b), allow us to find it in terms of $\hat{G}^{\mathrm{el}}$:
\begin{multline}\label{FelG}
\boldsymbol{\mathfrak{F}}^{\mathrm{el}}(\mathbf{q},i\omega_m) = \frac{g_\mathrm{v}T}{S} \sum_{\mathbf{p}, i\omega_{n}^{\prime}} \mathbf{p}\,\mathrm{Tr} \left[ \tau_{3}\hat{G}^{\mathrm{el}}(\mathbf{p},i\omega_{n}^{\prime})\right. \\
\times\left.\hat{G}^{\mathrm{el}}(\mathbf{p},i\omega_{n}^{\prime})\tau_{3} \hat{G}^{\mathrm{el}}(\mathbf{p+q},i\omega_{n}^{\prime}+i\omega_{m})  \right]\\
+(\mathbf{q},i\omega_{m} \to-\mathbf{q},-i\omega_{m}),
\end{multline}
where $\tau_{3}$ is the Pauli matrix and $i\omega_{n}^{\prime}=2\pi i(n+1/2)T$ are fermionic Matsubara frequencies. After substituting (\ref{GreenBCS}) into (\ref{FelG}) and performing summation over $i\omega_{n}^{\prime}$, we obtain the rectification function at nonzero temperature, see the full formula (\ref{26}) in Appendix \ref{Appendix_B}. As demonstrated in Fig. \ref{fig:polT} below, the temperature dependence of the drag density $\rho_\mathrm{dr}$ is weak in the vicinity of $T=0$. Therefore we can take the $T=0$ limit in our calculations without a significant error. In this limit the expression (\ref{26}) takes the simple form
\begin{multline}\label{Fel}
\boldsymbol{\mathfrak{F}}^{\mathrm{el}} (\mathbf{q},i\omega_m)\\
=\frac{8g_\mathrm{v}}{S}\sum_{\mathbf{p}}\frac{\mathbf{p} \left(E_{\mathbf{p}}^{\mathrm{el}}+E_{\mathbf{p+q}}^{\mathrm{el}}\right)i\omega_m L^{\mathrm{el}}(\mathbf{p},\mathbf{q})}{\left[(i\omega_m)^{2}-\left(E_{\mathbf{p}}^{\mathrm{el}}+E_{\mathbf{p+q}}^{\mathrm{el}}\right)^2\right]^{2}}.
\end{multline}
Here the fermionic coherence factor is $L^{\mathrm{el}}(\mathbf{p},\mathbf{q})= (u_{\mathbf{p}}v_{\mathbf{p+q}} + u_{\mathbf{p+q}}v_{\mathbf{p}})^{2}$; the same factor appears in the expressions for a polarization function of superconductor \cite{Anderson1958,Rickayzen1959,Gabovich1973}.

The major contribution to $\chi^{\mathrm{el-p}}_{ik}$ (\ref{chidiag}) is given by the frequencies around $\Omega_{\mathrm{R}}$, since the Bogoliubov excitation energy is $E^{\mathrm{p}}_{\mathbf{q}} \approx \Omega_{\mathrm{R}}$ at the characteristic interlayer momentum transfer $q \sim 1/L$ \cite{Narozhny2016}. Having it in mind, one can find that in the limit $T=0$, for a weak-coupling superconductor ($\Delta\ll E_\mathrm{F}$ and $\Delta\ll\Omega_{\mathrm{R}}$), the expression (\ref{Fel}) can be approximated by
\begin{multline}\label{Fel0}
\boldsymbol{\mathfrak{F}}^{\mathrm{el}} (\mathbf{q},i\omega_m) \\ \approx \frac{4ig_\mathrm{v}}{S}\mathrm{Im}\sum_{\mathbf{p}}\frac{\mathbf{p}\,\delta(\epsilon^{\mathrm{el}}_{\mathbf{p}})}{i\omega_{m}-\mathbf{pq}/m^{*}_{\mathrm{e}}-q^{2}/2m^{*}_{\mathrm{e}}},
\end{multline}
where $\delta(x)$ is the Dirac delta function.

\subsection{Interaction}
\subsubsection{Screening}

Taking into account the screening of the interlayer electron-polariton interaction (\ref{VepBorn}) is crucial for accurate quantitative description of ABE in a system containing the metallic-like 2DEG layer. The screened electron-polariton interaction in the two-layer system \cite{Narozhny2016}
\begin{equation}\label{Vep0}
V^{\mathrm{el-p}}_{\mathrm{scr}}(q,i\omega_m)=\frac{V^{\mathrm{el-p}}_{\mathbf{k}=0}(q)}{\varepsilon(q,i\omega_m)} = X_{0}X_{q}\frac{V^{\mathrm{el-x}}(q)}{\varepsilon(q,i\omega_m)}
\end{equation}
is written in terms of the bare electron-exciton interaction $V^{\mathrm{el-x}}(q)$ (which is considered in Sec.~\ref{sec:Vind} and Sec.~\ref{sec:Vdir} below) and the dielectric function
\begin{multline}\label{epsilon0}
\varepsilon(q,i\omega_m)\\ = \{1-V^{\mathrm{el-el}}(q)\Pi^{\mathrm{el}}(q,i\omega_m)\}\{1-g^{\mathrm{p-p}}_{0\mathbf{q}}(\mathbf{q})\Pi^{\mathrm{p}}(q,i\omega_m)\} \\ - [V^{\mathrm{el-p}}_{0}(q)]^{2} \Pi^{\mathrm{el}}(q,i\omega_m) \Pi^{\mathrm{p}}(q,i\omega_m).
\end{multline}
Here $\Pi^{\mathrm{el}}(q,i\omega_m)$ is the polarization, or density response, function of the superconducting 2DEG, and $\Pi^{\mathrm{p}}(q,i\omega)$ is the polarization function of the noninteracting polariton system; $V^{\mathrm{el-el}}(q) = 2\pi e^{2}/q\varepsilon_{\mathrm{env}}$ is the bare Coulomb interaction between electrons, $\varepsilon_{\mathrm{env}}$ is the mean dielectric constant of a medium surrounding exciton and electron layers. The main contribution to  $\Pi^{\mathrm{p}}(q,i\omega_m)$ is given by the processes involving the condensate \cite{Griffin,Tanatar2020} and can be written as $\Pi^{\mathrm{p}}(q,i\omega_m)=2n_{0}^{\mathrm{p}}\tilde\epsilon^{\mathrm{p}}_{\mathbf{q}}/[(i\omega_m)^{2}-(\tilde\epsilon^{\mathrm{p}}_{\mathbf{q}})^{2}]$.

The Green functions of polaritons (\ref{GreenPolariton}), which are linked to the interlayer interaction lines in Fig.~\ref{fig: Diagram for EP}, already include the interaction screening in the excitonic layer via the density response function $\Pi^{\mathrm{p}}(q,i\omega_m)$. Therefore using $V^{\mathrm{el-p}}_{\mathrm{scr}}$ given by the formulas (\ref{Vep0})--(\ref{epsilon0}) in the final expression (\ref{chidiag}) for the current response would lead to double counting of the diagrams involving the power series of $g^{\mathrm{p-p}}_{0\mathbf{q}}(\mathbf{q})\Pi^{\mathrm{p}}(q,i\omega_m)$. To avoid this double counting, we use in Eq.~(\ref{chidiag}), instead of the fully screened interaction (\ref{Vep0}), the partially screened interlayer interaction $\tilde{V}^{\mathrm{el-p}}_{\mathrm{scr}}(q,i\omega_m) = V^{\mathrm{el-p}}_{\mathrm{scr}}(q,i\omega_m)[1-g^{\mathrm{p-p}}_{0\mathbf{q}}(\mathbf{q})\Pi^{\mathrm{p}}(q,i\omega_m)]$. It can be written as
\begin{equation}\label{TildeVel-p}
\tilde{V}^{\mathrm{el-p}}_{\mathrm{scr}}(q,i\omega_m)=\frac{V^{\mathrm{el-p}}_{\mathbf{k}=0}(q)}{\tilde{\varepsilon}(q,i\omega_m)}=X_0X_q\frac{V^{\mathrm{el-x}}(q)}{\tilde{\varepsilon}(q,i\omega_m)}
\end{equation}
in terms of
\begin{multline}\label{epsilon}
\tilde{\varepsilon}(q,i\omega_m) = 1-V^{\mathrm{el-el}}(q)\Pi^{\mathrm{el}}(q,i\omega_m)\\-[V^{\mathrm{el-p}}_{0}(q)]^{2} \Pi^{\mathrm{el}}(q,i\omega_m) \tilde\Pi^{\mathrm{p}}(q,i\omega_m).
\end{multline}
Here the polarization function of interacting polaritons is
\begin{multline}
\tilde\Pi^{\mathrm{p}}(q,i\omega_m)= \frac{ \Pi^{\mathrm{p}}(q,i\omega_m)}{1-g^{\mathrm{p-p}}_{0\mathbf{q}}(\mathbf{q})\Pi^{\mathrm{p}}(q,i\omega_m)}\\=n_{0}^{\mathrm{p}} \left[  G_{11}^{\mathrm{p}} (\mathbf{q},i\omega_{m}) + G_{12}^{\mathrm{p}} (\mathbf{q},i\omega_{m}) + \ G_{21}^{\mathrm{p}} (\mathbf{q},i\omega_{m})\right. \\ \left. + G_{22}^{\mathrm{p}} (\mathbf{q},i\omega_{m})  \right]  = \frac{2 n_{0}^{\mathrm{p}} \tilde{\epsilon}^{\mathrm{p}}_{\mathbf{q}}}{(i\omega_{m})^{2}-\left(E^{\mathrm{p}}_{\mathbf{q}} \right)^{2}}.
\end{multline}

For $\Pi^{\mathrm{el}}(q,i\omega_m)$ we use the polarization function of normal 2DEG, instead of that of a superconductor \cite{Gabovich1973, Rickayzen1959, Anderson1958}, because these functions are almost equal in the range of momenta and frequencies $\omega \sim \Omega_{\mathrm{R}}, q\sim L^{-1}$, which provide the dominating contribution to Eq.~(\ref{chidiag}). The polarization function of normal 2DEG can be evaluated in the random phase approximation (RPA) analytically \cite{Stern1967}:
\begin{multline}\label{Pol}
\Pi^{\mathrm{el}}_{\mathrm{RPA}}(q,i\omega_m) = -2g_\mathrm{v}\vartheta \frac{p_{\mathrm{F}}}{q} \left\{
\vphantom{\sqrt{1-\left(\frac{q}{2p_{\mathrm{F}}}-i\frac{\omega}{qv_{\mathrm{F}}}\right)^{2}}}\frac{q}{2p_{\mathrm{F}}}+i\,\mathrm{sign}(\omega_m)\right.\\ \left.\times  \sqrt{1-\left(\frac{q}{2p_{\mathrm{F}}}-\frac{i\omega_m}{qv_{\mathrm{F}}}\right)^{2}} \right\} + \mbox{c.c.},
\end{multline}
where $p_{\mathrm{F}}, v_{\mathrm{F}}$ are the Fermi momentum and velocity, $\vartheta = m_{\mathrm{e}}^{*}/2\pi$ is the density of states at the Fermi level of 2DEG.

In contrast to our approach which takes into account the dynamical screening, in the previous works on polariton-mediated superconductivity and normal electron-polariton drag \cite{Boev2019,Laussy2010, Laussy2012, Cotlet2016}, the Thomas-Fermi (TF) static long-wavelength approximation for the 2DEG polarization function was used: $\Pi^{\mathrm{el}}_{\mathrm{TF}}(q,i\omega_m)=-2g_\mathrm{v}\vartheta$. To compare the results obtained in various approximations, we perform calculations using both $\Pi^{\mathrm{el}}_{\mathrm{TF}}$ and $\Pi^{\mathrm{el}}_{\mathrm{RPA}}$. As will be demonstrated below in Fig. \ref{fig:RPA}, the magnitude of the predicted ABE depends significantly on the type of screening.

\subsubsection{Indirect excitons}\label{sec:Vind}

In the system of electrons and indirect dipolar excitons, depicted in Fig.~\ref{fig:ExpSet}(a), the bare electron-exciton interaction is \cite{Laussy2010}
\begin{multline}\label{Vxind}
V_{\mathrm{ind}}^{\mathrm{el-x}}(q)=\frac{2 \pi e^{2}}{\varepsilon_{\mathrm{env}} q}\left\{\frac{e^{-q\left(L-\beta_{\mathrm{e}} d\right)}}{\left[1+\left(\frac12\beta_{\mathrm{h}} q a_{\mathrm{B}}\right)^{2}\right]^{3 / 2}} \right. \\ \left. -\frac{e^{-q\left(L+\beta_{\mathrm{h}} d\right)}}{\left[1+\left(\frac12\beta_{\mathrm{e}} q a_{\mathrm{B}}\right)^{2}\right]^{3 / 2}}\right\}.
\end{multline}
Here $\beta_{\mathrm{e,h}} = m_{\mathrm{e,h}}/(m_{\mathrm{e}}+m_{\mathrm{h}})$, $a_{\mathrm{B}}$ is the Bohr radius of exciton, $L$ is the mean distance between electrons and excitons in the out-of-plane direction, $d$ is the distance between the electron and hole layers, involved into exciton formation. This expression is obtained under the assumption that the exciton has the 1s 2D hydrogen-like wave function in the in-plane directions, which is not perturbed by the electron-hole interlayer separation and by the 2DEG layer. More complicated forms of the exciton wave function were studied, for example, in Refs.~\cite{Leavitt1990, Gribakin2021}.

\subsubsection{Direct excitons}\label{sec:Vdir}

In the case of direct excitons, shown in Fig.~\ref{fig:ExpSet}(b), the interlayer interaction is qualitatively different since excitons do not possess a constant dipole moment. The exciton with an in-plane polarizability $\alpha$ affected by the in-plane electrostatic field $E_{||}$ of the electron acquires the energy shift $-\frac12\alpha E_{||}^{2}$. Therefore the electron-exciton interaction energy at the in-plane distance $r$ between electron and exciton is
\begin{equation}\label{Vxdir}
V_{\mathrm{dir}}^{\mathrm{el-x}}(r) =  - \frac{e^{2}\alpha}{2\varepsilon_{\mathrm{env}}^{2}}\frac{r^{2}}{(r^{2}+L^{2})^{3}}.
\end{equation}
The Fourier transform of this expression is
\begin{equation}
V_{\mathrm{dir}}^{\mathrm{el-x}}(q) = - \frac{e^{2}\alpha}{2\varepsilon_{\mathrm{env}}^{2}} \frac{\pi q}{L} \left\{ \frac{1}{2}K_{1}(qL) - \frac{qL}{4} K_{0}(qL) \right\},
\end{equation}
where $K_{i}$ are modified Bessel functions of the second kind. We disregard the out-of-plane polarizability of the exciton, because in 2D materials it is 3-4 orders of magnitude lower than $\alpha$ \cite{Garm}.

The polarizability of a 2D hydrogen-like exciton $\alpha=21 \varepsilon a_{\mathrm{B}}^{3}/128$ with the Bohr radius $a_{\mathrm{B}}=0.5$ nm in TMDC is $\alpha\approx0.1$ nm$^{3}$. However, for numerical calculations we take much larger value $\alpha = 30$ nm$^{3}$, which was evaluated in Ref.~\cite{Garm} with accounting for the Rytova-Keldysh screening of interaction between electron and hole, forming the exciton: the screening by the TMDC layer makes the exciton more weakly bound and thus more sensitive to perturbations, so its polarizability is larger than in the 2D hydrogen model.

\begin{table}[t]
\centering
\begin{tabular}{|l||l|l|}
\hline
Parameter & TMDC & QW \\
\hline
\multicolumn{3}{|c|}{Common parameters} \\
\hline
$m_{\mathrm{e}},m_{\mathrm{h}}$ & 0.5$m_{0}$, 0.5$m_{0}$  & 0.067$m_{0}$, 0.45$m_{0}$   \\
$m_{\mathrm{e}}^{*}$ & 0.5$m_{0}$ & 0.067 $m_{0}$ \\
$m_{\mathrm{c}}$ & $5 \times 10^{-5} m_{0}$ & $5 \times 10^{-5} m_{0}$\\
$n^{\mathrm{el}}$ & $10^{13}$ cm$^{-2}$ & $10^{12}$ cm$^{-2}$ \\
$n^{\mathrm{p}}_{0}$ & $10^{12}$ cm$^{-2}$ & $10^{10}$ cm$^{-2}$  \\
$\varepsilon_{\mathrm{env}}$ & 7 & 7\\
\hline
\multicolumn{3}{|c|}{Parameters for indirect excitons} \\
\hline
$d$ & 4 nm & 10 nm \\
$L$ & 15 -- 100 nm & 15 -- 100 nm \\
$\Omega_{\mathrm{R}}$  &    10 -- 40 meV  & 10 -- 40 meV \\
$a_{\mathrm{B}}$ & 0.5 nm &  17 nm \\
$g^{\mathrm{x-x}}$ & 0.1 $\mathrm{\mu eV}\cdot \mathrm{\mu m}^{2}$ &  1.3 $\mathrm{\mu eV}\cdot \mathrm{\mu m}^{2}$ \\
\hline
\multicolumn{3}{|c|}{Parameters for direct excitons} \\
\hline
$\alpha$ & 30 nm$^{3}$&  --- \\
$L$ & 4 -- 15 nm & --- \\
$\Omega_{\mathrm{R}}$  &    30 -- 100 meV  & --- \\
$g^{\mathrm{x-x}}$ & 0.1 $\mathrm{\mu eV\cdot\mu m}^{2}$&  --- \\
\hline
\end{tabular}
\caption{Parameters for our calculations; $n^{\mathrm{el}}$ is the density of 2DEG, $m_{0}$ is the free electron mass. The polarizability $\alpha$ for TMDC is taken from Ref.~\cite{Garm} at $\varepsilon_{\mathrm{env}} = 7$. Other parameters are taken from \cite{Laussy2012, QFl,Sven2021,Estrecho2019,Kasprzak2006,Zhao2021,Cotlet2016,HuiHu2020}.}
\label{table:1}
\end{table}

\section{Results}\label{Sec3}
\subsection{Drag density} \label{dragdensity}

The most widely used materials for exciton polaritons in optical microcavity are semiconductor QWs based on GaAs or CdTe heterostructures and, recently, TMDC layers \cite{Basov2020}. For our numerical calculations we take the typical parameters of these materials \cite{QFl,Sven2021,Estrecho2019,Kasprzak2006,Zhao2021}, which are listed in Table \ref{table:1}. The density of polariton condensate $n_0^\mathrm{p}$ is assumed to be 1-2 orders of magnitude lower than the Mott density, which can be approximated as $n_{\mathrm{M}} \sim 1/a_{\mathrm{B}}^{2}$ \cite{Sven2021,Estrecho2019}.

For the polariton-polariton interaction, we use the expression (\ref{gppBorn}), which relates it to the exciton-exciton interaction constant $g^{\mathrm{x-x}}$. This formula is widely used in the literature and adequately describes the experimental results \cite{Estrecho2019}, despite that questions about its validity have been raised recently \cite{HuiHu2020}. For $g^{\mathrm{x-x}}$ in the case of indirect excitons, we will use the formula derived in Ref. \cite{dipdip}:
\begin{equation}\label{gxx}
g^{\mathrm{x-x}} = \frac{e^{2}a_{\mathrm{B}}}{4 \pi\varepsilon_{\mathrm{env}}} \left(  6+3.5 \frac{d}{a_{\mathrm{B}}}  \right).
\end{equation}
It gives $g^{\mathrm{x-x}}=1.3\, \mathrm{\mu eV\cdot\mu m}^{2}$ and $0.1\,\mathrm{\mu eV\cdot\mu m}^{2}$ for excitons in, respectively, QWs and TMDC.

The dielectric screening by the environment $\varepsilon_{\mathrm{env}}$ depends on the materials, which are used to fabricate the microcavity, electron and exciton layers. For instance, the dielectric constants for hexagonal boron nitride, TMDC, and GaAs are, respectively, about 4.5 \cite{epshBN}, 10 \cite{epshBN}, and 13 \cite{epsGaAs}. Moreover, since the interaction between indirect excitons and electrons (\ref{Vxind}) is inversely proportional to $\varepsilon_{\mathrm{env}}$ and squared in Eq.~(\ref{chidiag}), and the dielectric function in RPA (\ref{epsilon}) is slightly weakened by $\varepsilon_{\mathrm{env}}$, the superfluid drag density $\rho_\mathrm{dr}$ decreases with increasing $\varepsilon_{\mathrm{env}}$ approximately as $\rho_\mathrm{dr}\propto\varepsilon_{\mathrm{env}}^{-1}$. The same concerns the interaction between direct excitons and electrons (\ref{Vxdir}), if we account for the fact that $\alpha$ is roughly proportional to $\varepsilon_{\mathrm{env}}$ \cite{Garm}. The interaction between excitons (\ref{gxx}) also depends on $\varepsilon_{\mathrm{env}}$, but it does not change the general trend $\rho_\mathrm{dr}\propto\varepsilon_{\mathrm{env}}^{-1}$, since $g^{\mathrm{p-p}}$ essentially affects the dispersion of Bogoliubov excitations only at momenta much lower than the characteristic momentum $q\sim L^{-1}$ contributing to the integral in Eq.~(\ref{drag}). Therefore, similarly to \cite{Laussy2010,Laussy2012}, for the dielectric constant in our calculations we take the average value $\varepsilon_{\mathrm{env}}=7$.

\begin{figure}[t]
\centering
\includegraphics[width=\columnwidth]{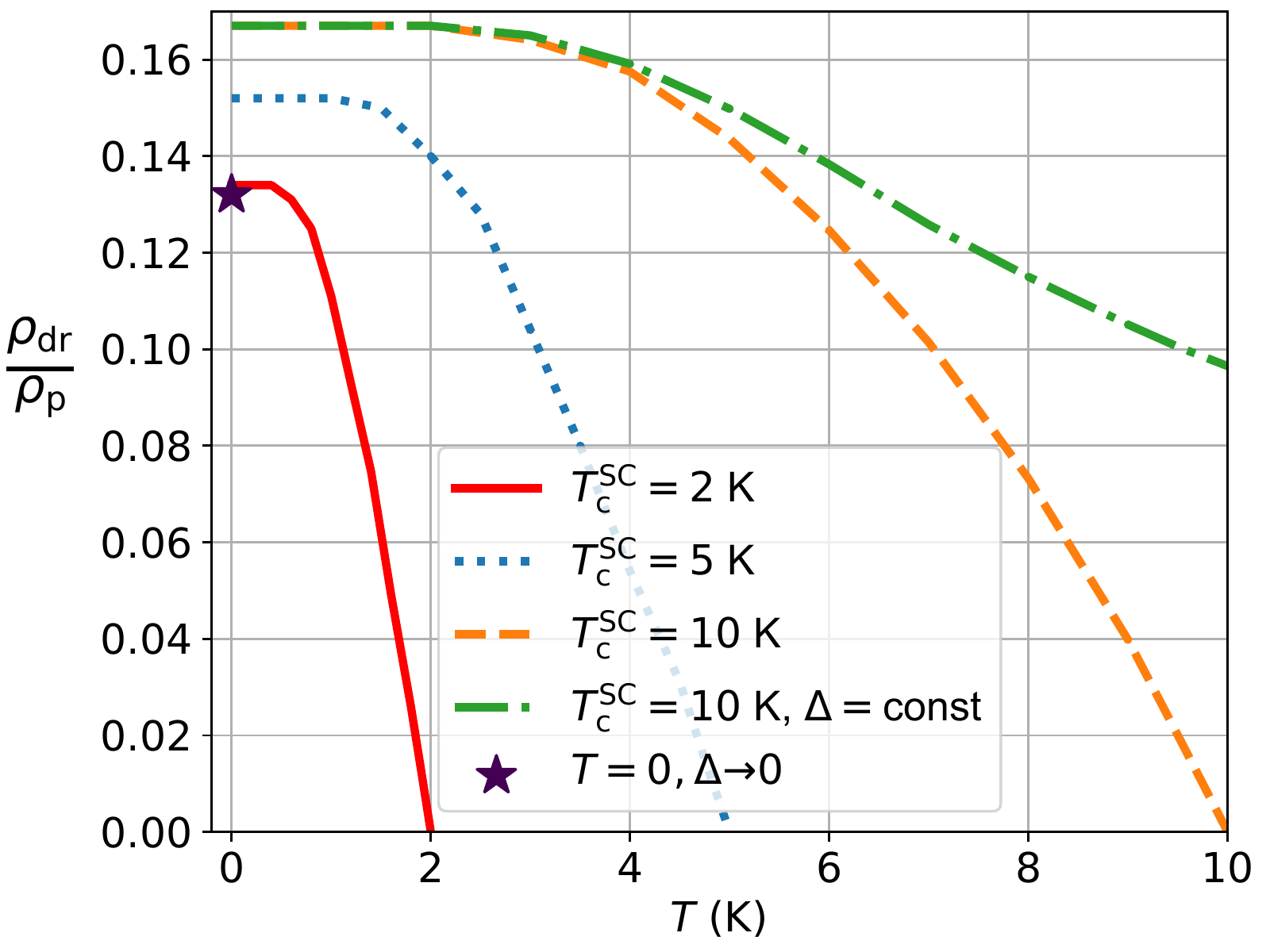}
\caption{Drag density $\rho_\mathrm{dr}$ as a function of temperature for the superconductor critical temperatures $T^{\mathrm{SC}}_{\mathrm{c}}=2$ K (red solid line), 5 K (blue dotted line) and 10 K (orange dashed line). Green dash-dotted line shows $\rho_\mathrm{dr}$ calculated with the energy gap independent of the temperature: $\Delta=17.6\,\mbox{K}$ for the case $T^{\mathrm{SC}}_{\mathrm{c}}=10\,\mbox{K}$; the star is the calculation result with the approximate formula (\ref{Fel0}) for the superconductor rectification function in the $T=0$, $\Delta\to 0$ limit. The system parameters are the same as in the Fig. \ref{fig:pol3L}(b) for the case of $L=20$~nm, $\Omega_{\mathrm{R}}=10$ meV, when both 2DEG and excitonic layers are based on semiconductor QWs.}
\label{fig:polT}
\end{figure}

In the following we relate the superfluid drag density $\rho_\mathrm{dr}$ to the total mass density of the polariton gas $\rho_{\mathrm{p}}=m_{\mathrm{p}} n^{\mathrm{p}}\approx m_{\mathrm{p}} n^{\mathrm{p}}_{0}$, which is dominated by the condensate in the weakly-interacting regime.

First, we reassure that $\rho_{\mathrm{dr}}$ does not significantly depend on the temperature, and we can use the approximate expression (\ref{Fel0}) for $T=0$ instead of the full formula (\ref{26}). The typical temperature dependence of $\rho_{\mathrm{dr}}$ is depicted in Fig.~\ref{fig:polT}. While at low temperatures $\rho_{\mathrm{dr}}$ tends to a constant, it vanishes linearly as $1-T/T^{\mathrm{SC}}_{\mathrm{c}}$ when $T$ approaches the critical temperature $T^\mathrm{SC}_\mathrm{c}$ of the superconductor. We assumed that the critical temperature  $T^{\mathrm{BEC}}_{\mathrm{c}}$ of polaritonic BEC is much higher than $T^{\mathrm{SC}}_{\mathrm{c}}$, and the condensate thermal depletion is negligible. Fig.~\ref{fig:polT} demonstrates that the behavior of $\rho_{\mathrm{dr}}(T)$ is determined mainly by the temperature dependence of the superconducting energy gap $\Delta(T)$. Therefore, at low enough temperature $T\ll\Delta$, we can take the $T\to 0$ limit in the calculations. Moreover, at $T\to 0$ the drag density weakly depends on $\Delta(0)$, as also seen in Fig.~\ref{fig:polT}, since $\Delta(0)$ is the smallest energy scale in the system at weak coupling. Thus we will use Eq.~(\ref{Fel0}) for the rectification function of 2DEG in the limit $T=0$, $\Delta\rightarrow0$, which provides the lower bound for $\rho_\mathrm{dr}$ at $T\ll T^\mathrm{SC}_\mathrm{c}\ll T^{\mathrm{BEC}}_{\mathrm{c}}$.

\begin{figure}[t]
\centering
\includegraphics[width=\columnwidth]{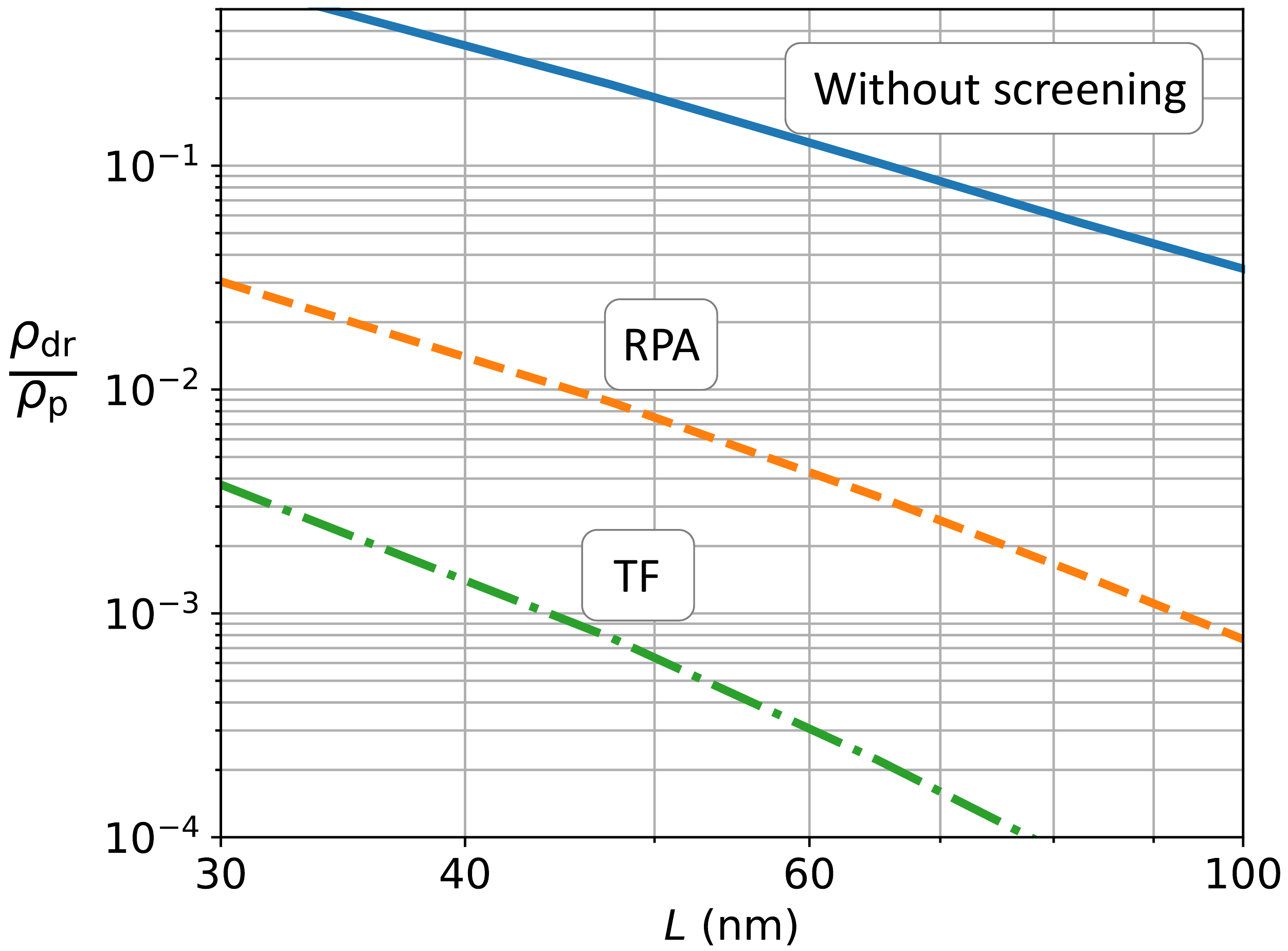}
\caption{Drag density $\rho_\mathrm{dr}$ as a function of interlayer distance $L$ for different kinds of interaction: unscreened and screened in RPA and in TF approximation. Parameters are the same as in Fig.~\ref{fig:pol3L}(b) for the case of $\Omega_{\mathrm{R}}=10$ meV, when both 2DEG and excitonic layers are based on QWs.}
\label{fig:RPA}
\end{figure}

In Fig.~\ref{fig:RPA} the drag densities, calculated without the screening of interaction by both electrons and polaritons ($\Pi^{\mathrm{p}}=\Pi^{\mathrm{el}}=0$), and with the screening in RPA and in TF approximation are presented. In the TF approximation, $\rho_{\mathrm{dr}}$ is by an order of magnitude lower than in RPA, because the interaction screening is overestimated in the static TF limit. On the other hand, the calculation without the screening provides $\rho_{\mathrm{dr}}$ an order of magnitude higher than the RPA result because of the overestimated interlayer interaction. Later in the calculations we will use RPA, which adequately describes both the screening and the dynamical effects.

\begin{figure}[!t]
\centering
\includegraphics[width=\columnwidth]{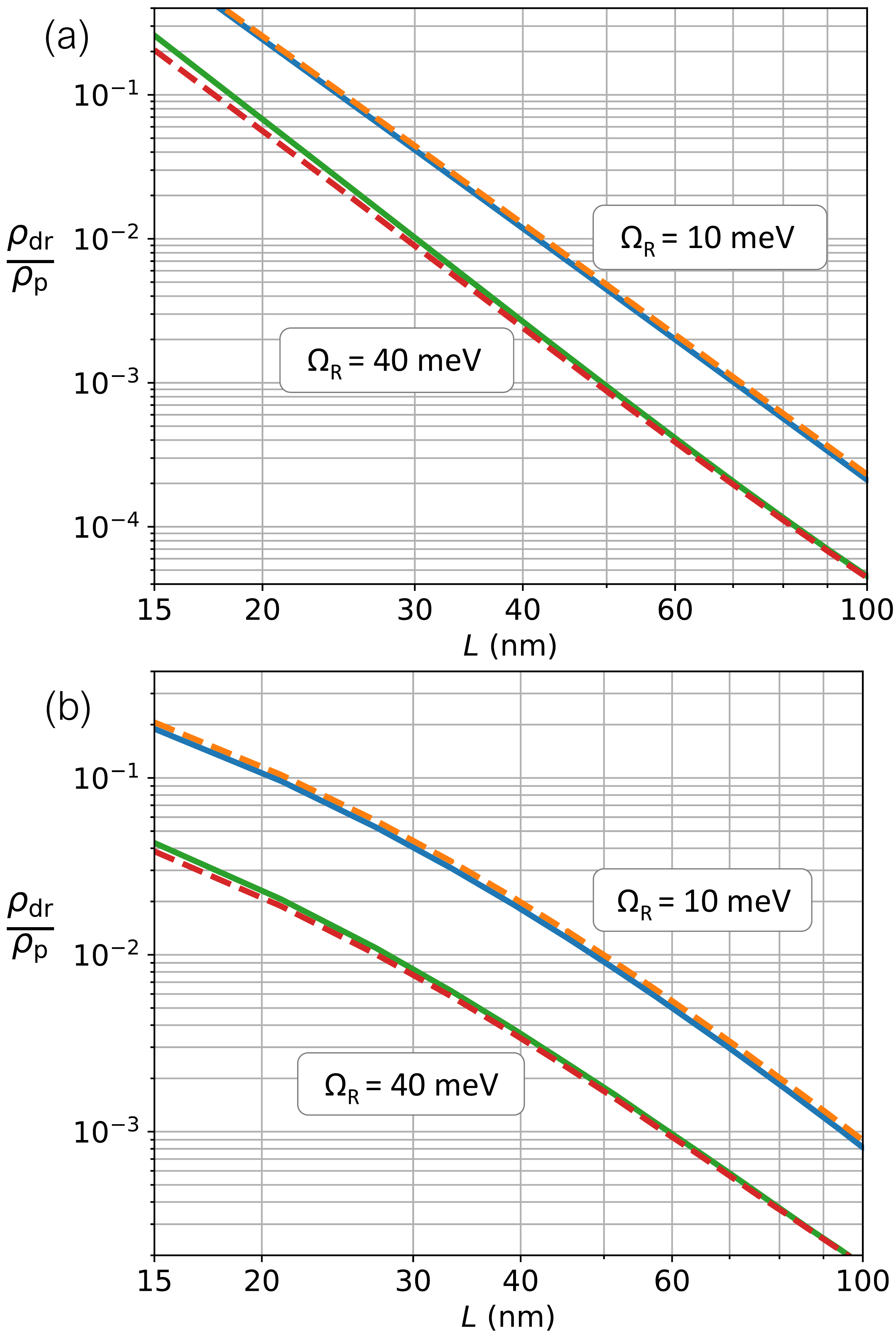}
\caption{Drag density $\rho_\mathrm{dr}$ in systems with indirect polaritons whose excitons are located in TMDC (a) and QWs (b) with the Rabi splittings ${\Omega_\mathrm{R}=10 \,\mbox{meV}}$ and 40 meV. Solid and dashed lines show the cases of electronic layers based, respectively, on TMDC and QW. The polariton subsystem parameters for the upper panel (TMDC) are: ${n_{0}^\mathrm{p} = 10^{12}}$~cm$^{-2}$, $m_{\mathrm{e}}=m_{\mathrm{h}}=0.5m_{0}$, $d=4$ nm, $a_{\mathrm{B}}=0.5\,\mbox{nm}$, ${g^{\mathrm{x-x}} =0.1}$~$\mathrm{ \mu eV}\cdot \mathrm{\mu m}^{2}$; for the lower panel (QWs): $n_{0}^{\mathrm{p}} = 10^{10}$~cm$^{-2}$, $m_{\mathrm{e}}=0.067m_{0}, m_{\mathrm{h}}=0.45m_{0}$, $d=10$ nm, $a_{\mathrm{B}}=17\,\mbox{nm}$, $g^{\mathrm{x-x}} =1.3$~$\mathrm{\mu eV}\cdot \mathrm{\mu m}^{2}$. The parameters of the TMDC electronic layer (solid lines): $m^{*}_{\mathrm{e}}=0.5m_{0}, n^{\mathrm{el}}=10^{13}\,\mbox{cm}^{-2}$; for the QW electronic layer (dashed lines): $m^{*}_{\mathrm{e}}=0.067m_{0}, n^{\mathrm{el}}=10^{12}$ cm$^{-2}$.}  \label{fig:pol3L}
\end{figure}

\begin{figure}[t]
\centering
\includegraphics[width=\columnwidth]{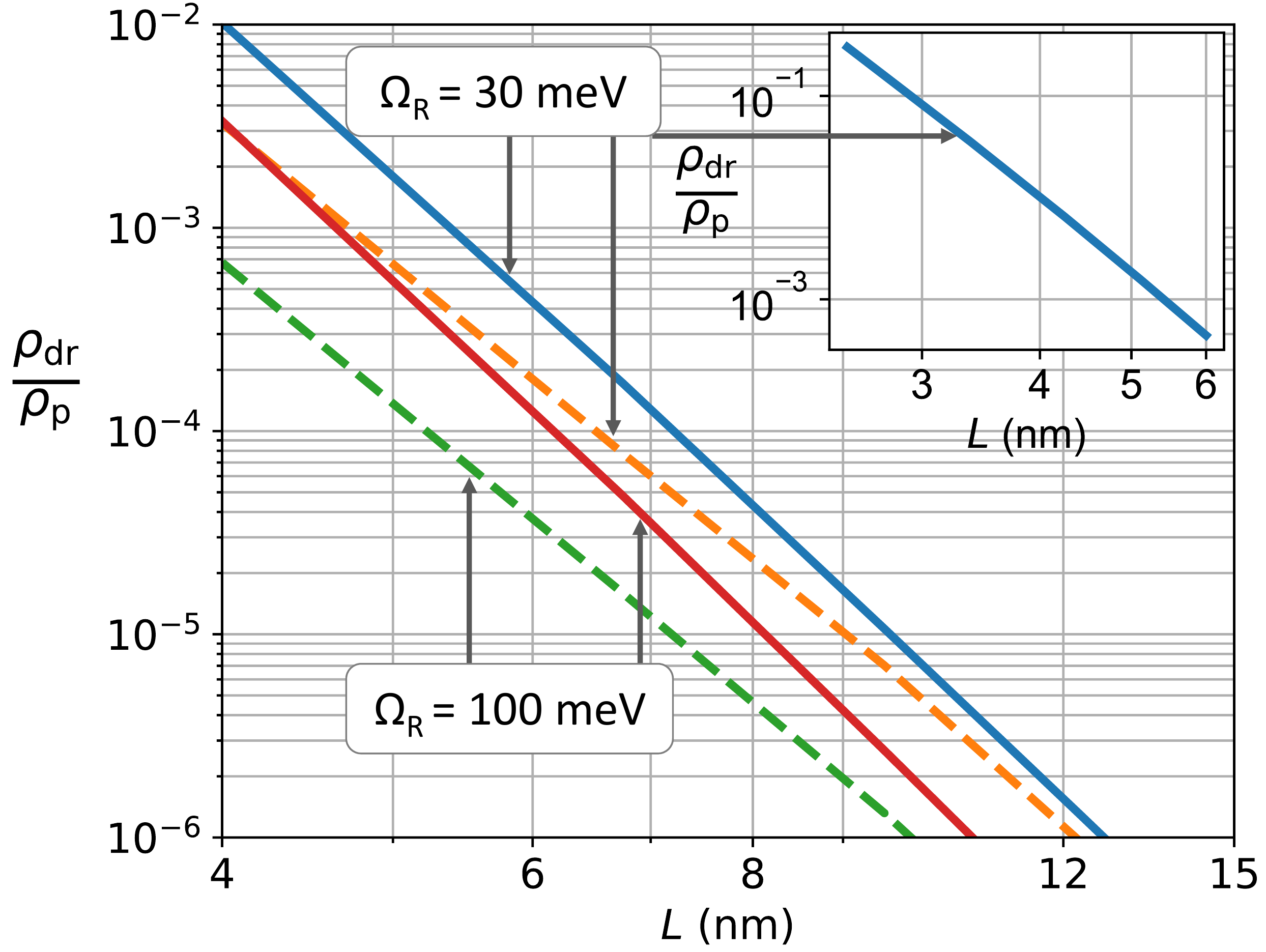}
\caption{Drag density $\rho_\mathrm{dr}$ in the system of direct polaritons with the excitons in TMDC layer at $\Omega_{\mathrm{R}}=30$ meV and 100~meV. Solid and dashed lines correspond to TMDC and QW electron layers. Inset is for the case when $\Omega_{\mathrm{R}}=30\,\mbox{meV}$, and both electrons and polaritons are in TMDC, so the electron and exciton layers can be put very close to each other. The system parameters are the same as in Fig. \ref{fig:pol3L}(a) with the exciton polarizability $\alpha = 30$ nm$^{3}$.}
\label{fig:pol2L}
\end{figure}

Figs. \ref{fig:pol3L} and \ref{fig:pol2L} show the dependence of $\rho_{\mathrm{dr}}$ on the distance between excitonic and electronic layers for indirect and direct polaritons respectively. In the case of indirect polaritons (Fig.~\ref{fig:pol3L}), we consider the realizations of both exciton and electron layers on the base of QWs and TMDCs. In the case of direct polaritons (Fig.~\ref{fig:pol2L}) we do not consider the excitons in QW, because in such systems multiple QW structures are used to make Rabi splitting higher, and therefore the mean distance between excitons and electrons would be of the order of 100 nm \cite{Estrecho2019}, when the drag effect is too weak.

We note that $\rho_\mathrm{dr}$ markedly decreases with increase of the Rabi splitting $\Omega_\mathrm{R}$ because the latter provides the energy scale for virtual lower polariton excitations at the characteristic interlayer momentum transfer $q\sim L^{-1}$; these excitations are responsible for the drag so increase of their energy decreases $\rho_\mathrm{dr}$. Realization of indirect polaritons on the base of TMDC (Fig. \ref{fig:pol3L}(a)) is more preferable for observation of ABE, because in this case $L$ can be made smaller in practice, although $\rho_\mathrm{dr}$ is almost the same at equal values of $L$ for both realizations of indirect polaritons, as seen from comparison of Fig.~\ref{fig:pol3L}(a) and Fig.~\ref{fig:pol3L}(b). The specific realization of 2DEG are not important for the case of indirect polaritons, as seen in Fig. \ref{fig:pol3L} from comparison of solid and dashed lines. In the case of direct polaritons in Fig. \ref{fig:pol2L}, $\rho_\mathrm{dr}$ is generally lower than for indirect polaritons because of weaker electron-exciton interaction. Also $\rho_\mathrm{dr}$ in this case is higher when both 2DEG and excitons are located in TMDC layers (solid lines in Fig.~\ref{fig:pol2L}).

The drag density decreases with interlayer distance as $L^{-\beta}$, where $\beta\approx 4-5$ and $\beta\approx 2-4$ for the indirect polaritons in TMDC and QW respectively, and $\beta\approx 7-8$ for direct polaritons in TMDC. Although in the case of direct excitons (Fig.~\ref{fig:pol2L}) $\rho_{\mathrm{dr}}$ decreases faster with $L$ than in the systems with indirect excitons (Fig.~\ref{fig:pol3L}), the effect in the former case could be still appreciable at the distances $L\sim (2-4)$ nm, since TMDC layers can be put very close to each other due to their extreme thinness (see the inset in Fig.~\ref{fig:pol2L}). For a strong intercomponent interaction the drag saturates, and the system should exhibit transition from the double-superfluid phase to a paired superfluid phase \cite{Nespolo2018, Selin2018}. Our theory is not aimed to describe this regime, so we limit our calculations to $\rho_{\mathrm{dr}}<0.5\rho_\mathrm{p}$.

\begin{figure}[t]
\centering
\includegraphics[width=\columnwidth]{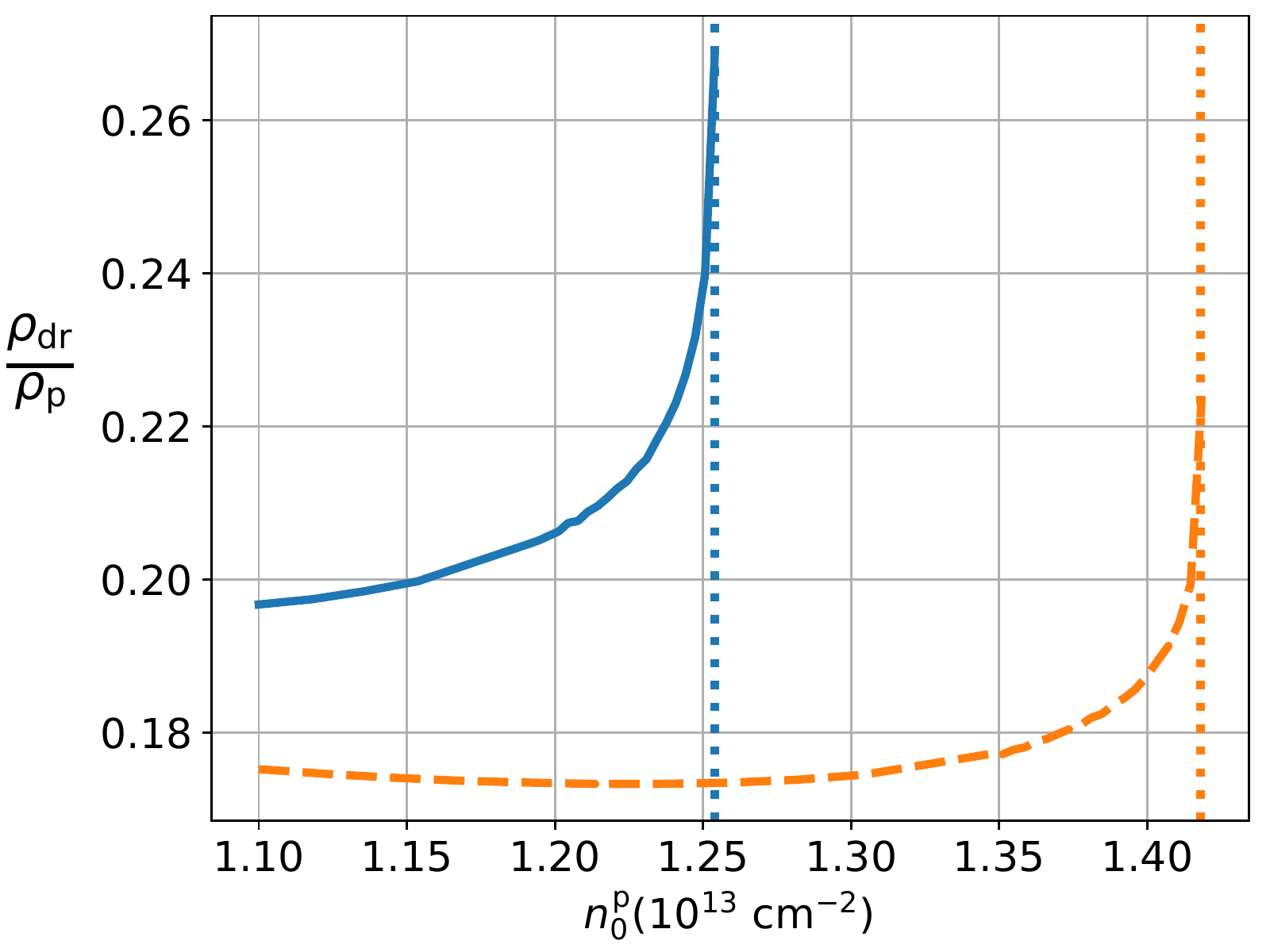}
\caption{Drag density $\rho_\mathrm{dr}$ as a function of polariton density $n_0^\mathrm{p}$ in the system with TMDC-based indirect polaritons. Solid and dashed lines correspond to, respectively, TMDC and QW electron layers. The system parameters are the same as in Fig.~5(a) for the case $\Omega_{\mathrm{R}} = 10\,\mbox{meV}$, $L = 20\,\mbox{nm}$. Vertical dotted lines show critical polariton densities where the roton minimum reaches zero energy, and the drag density diverges.}
\label{fig:roton}
\end{figure}

Finally, we consider the ABE in the strong-correlation regime described in \cite{Cotlet2016}. Due to 2DEG-induced screening of polariton-polariton interaction, dispersion of Bogoliubov excitations in the polariton system acquires a roton minimum, which becomes deeper as the polariton density $n_0^\mathrm{p}$ or electron-exciton interaction $V^{\mathrm{el-x}}$ increases. This dispersion is given by zeros of the dielectric function (\ref{epsilon0}). When the roton minimum reaches zero energy at some critical polariton density, the system becomes unstable against transition into a supersolid or other symmetry-breaking phase.

Since we screen the interlayer interaction (\ref{TildeVel-p}) by density responses of both polaritons and 2DEG, these strong-correlation effects are automatically taken into account in our approach in RPA. Fig.~\ref{fig:roton} shows behavior of the drag density $\rho_\mathrm{dr}$ at high enough $n_0^\mathrm{p}$ in the vicinity of the roton instability. It can be seen that $\rho_\mathrm{dr}$ sharply increases and formally diverges at the critical density. Thus the correlation effects can enhance ABE near the point of instability due to large contribution of low-energy Bogoliubov excitations to $\rho_\mathrm{dr}$, however the issues with the stability of polaritonic system and robustness of its superfluidity can arise in this regime. In particular, the polariton density and other parameters like polariton-polariton interaction strength should be fine-tuned \cite{Cotlet2016} in order to bring the system very close to the instability point.

\subsection{Observation of the drag}

Now we consider possible methods to observe the superfluid drag, or ABE, between BEC of polaritons and superconductor. We suggest to create a flow of polariton Bose condensate via resonant pump by an inclined incident laser beam \cite{Lerario2017} and to measure the induced superconducting current in the 2DEG layer. The corresponding schematic experimental setup is depicted in Fig.~\ref{fig:ExpSet}. The velocity of polaritons in the moving condensate $v_{\mathrm{p}}=(\omega/m_\mathrm{p}c)\sin\theta$ can be varied via change of the incident angle $\theta$. Here $\omega$ is the cavity photon frequency, $c$ is the speed of light in vacuum. There are other possibilities to create a current of polariton BEC by applying stress \cite{Snoke2013}, magnetic and electric fields \cite{Chervy2020} or via the mean-field potential of exciton cloud \cite{Myers2021,Snoke2013}.

As the main signature of ABE, the motion of polariton condensate induces the superconducting current in the electronic subsystem with the mass density $g_{\mathrm{el}}=\rho_{\mathrm{dr}}v_{\mathrm{p}}$, which is equivalent to the electric current density
\begin{align}\label{elcurrents}
j_\mathrm{el}=g_\mathrm{el}e/m_\mathrm{e}^*\equiv J_{0}\kappa, \\ \label{J0}
J_{0} = m_\mathrm{p}n_0^\mathrm{p}v_\mathrm{p}e/m_\mathrm{e}^*,
\end{align}
where $\kappa=\rho_{\mathrm{dr}}/\rho_{\mathrm{p}}$ is of the order of 0.001--0.1 in realistic conditions, see Figs.~\ref{fig:pol3L}--\ref{fig:pol2L}. The quantity $J_{0}$ can be interpreted as an upper theoretical limit of the drag current, achieved at the perfect drag $\kappa=1$, $\rho_{\mathrm{dr}}=\rho_{\mathrm{p}}$. This expression allows us to estimate the maximal magnitude of the drag current: taking $v_{\mathrm{p}} = 10^{8}$ cm/s \cite{Lerario2017}, we get $J_{0}\approx 1.6$ A/m for the case when both electronic and excitonic layers are based on TMDC and $J_{0}\approx 0.12$ A/m for a QW-based system. Assuming that the thicknesses of TMDC and QW are, respectively, 1 nm and 10 nm, these currents correspond to three-dimensional current densities $\sim 10^{9}$ A/m$^{2}$ and $\sim 10^{7}$ A/m$^{2}$ for TMDC and QW cases respectively. Multiplying $J_{0}$ by the characteristic $\kappa=0.01$, we get superfluid drag currents $\sim 10^{7}$ A/m$^{2}$ and $\sim 10^{5}$ A/m$^{2}$, which are of the order of critical currents in some conventional superconductors \cite{Brun2017,Uchihashi2017}, hence they are high enough to be measurable.

For detection of the normal drag effect in electron-polariton system, the reverse method \cite{Berman2016} was proposed: to induce a current in electronic system and to measure the change in angle distribution of photons, escaped from microcavity. The superfluid drag effect can hardly be measured by this way, since electron Cooper pairs are much slower than polaritons. The mean velocity of the superconducting condensate is limited by a critical velocity of superconducting electrons $v_{\mathrm{c}}\sim \Delta/p_{\mathrm{F}}$ and in 2D materials it is of the order of $10^{5}$ cm/s \cite{Brun2017,Uchihashi2017}. Therefore the mass current density of dragged polaritons $g_\mathrm{p}=\rho_{\mathrm{dr}}v_{\mathrm{c}}$ corresponds to the nonzero angle $\theta=\kappa m_{\mathrm{p}}v_{\mathrm{c}}c/\omega$ of photon emission from the polariton BEC. For the photon energy $\omega=1.7\,\mbox{eV}$ it is $\theta\sim\kappa \times 0.003^\circ$. However, since photons are emitted from a finite area of the polariton cloud, there is a variance of the photon in-plane momentum due to the uncertainty principle. It results in the angular broadening $\Delta\theta=c/\omega K$, which is $\sim 6^\circ$ for the characteristic width of polariton cloud $K = 1\,\mu\mbox{m}$. This estimate of $\Delta \theta$ is in agreement, by an order of magnitude, with the observed angle distribution of out-flying photons about $\Delta \theta \approx 2 ^\circ$ \cite{Lerario2017}. As a result, the inclination angle of the photon emission $\theta$ due to ABE is expected to be much smaller than the uncertainty $\Delta \theta$ even at large $\kappa$.

There is another phenomenon, namely a kind of a photon drag effect \cite{PhDrag}, with the properties similar to those of ABE, which can emerge in the system we consider. Since the electronic layer absorbs a part of the incident light, the absorbed microcavity photons transfer a part of their momentum to Cooper pairs of the superconductor, so the induced supercurrent can, in principle, compete with current induced by ABE. Assuming that in the worst case the probability of a photon in microcavity to be absorbed by the electron layer is $\Theta\sim 0.1$, and that the whole momentum of the absorbed photons is transferred to Cooper pairs, the photon drag current would be $j_{\mathrm{ph}}= J_{0}\Theta$. Thus the ratio of the superfluid and photon drag currents in the worst case could achieve $j_{\mathrm{el}}/j_{\mathrm{ph}} = \kappa / \Theta \sim 1$. Nevertheless, it is hard to tell to what extent the photon drag effect is essential in this system, since to our knowledge this effect in superconductors was not yet studied.

In addition, our approach is applicable to the superfluid drag effect between electrons and indirect excitons in the absence of optical microcavity. BEC of indirect excitons was observed, for example, in \cite{High2012}, and the review of early experiments on excitonic BEC can be found in \cite{Snoke2002}. To apply our theory to exciton-electron system, we assume zero Rabi frequency $\Omega_\mathrm{R}=0$ and carry out the same calculations as in the Section \ref{dragdensity}.

We have obtained the following result: at the same system parameters, listed in Table~\ref{table:1}, the absolute values of $\rho_\mathrm{dr}$ in systems of electrons and indirect excitons is of the same order of magnitude as in systems with indirect polaritons. The formulas (\ref{elcurrents})--(\ref{J0}) for the drag-induced electron current density can be rewritten for the case of excitons as
\begin{equation}
j_{\mathrm{el}}=\rho_{\mathrm{dr}}v_{\mathrm{x}}e/m_{\mathrm{e}}^*,\label{j_el_ex}
\end{equation}
where $v_{\mathrm{x}}$ is the velocity of the exciton condensate. Since the exciton mass $m_\mathrm{x}$ is typically $10^4$ times larger than the polariton mass $m_\mathrm{p}\approx m_\mathrm{c}$, the maximal velocity of superfluid excitonic condensate $v_{\mathrm{x}}$, allowed by the Landau criterion as the sound velocity $\sqrt{g^{\mathrm{x-x}} n_{0}^\mathrm{x} /m_\mathrm{x}}$ \cite{Amo2009}, is two orders of magnitude lower than the maximal velocity of polaritons at the same exciton density $n_0^\mathrm{x}=n_0^\mathrm{p}$. Therefore, it can be concluded that in the case of excitons the superconducting drag-induced current (\ref{j_el_ex}) should be $\sim100$ times lower, so possible observation of ABE in this case is more challenging, although not excluded completely. Besides, since the ratio between the drag density and the full density of exciton condensate does not exceed $10^{-4}$, the superfluid drag effect cannot noticeably alter the vortex dynamics in the condensate and, accordingly, Berezinskii-Kosterlitz-Thouless transition \cite{Karle2019}.

\section{Conclusions}\label{Sec4}

The Andreev-Bashkin effect, or superfluid drag, was considered in the Bose-Fermi system consisting of Bose-condensed excitonic polaritons interacting with superconducting 2DEG. The excitonic part of polaritons as well as 2DEG are assumed to be located in parallel layers embedded into an optical microcavity. The main quantity characterizing this effect, the superfluid drag density $\rho_\mathrm{dr}$, was found from the current-current response function, which is calculated in the second order in the screened interlayer electron-exciton interaction.

Two observations about $\rho_\mathrm{dr}$ can be made: first, its decrease with increasing the temperature $T$ is mainly caused by the thermal suppression of the superconducting energy gap $\Delta(T)$, while at low enough temperature, $T\ll\Delta(0)$, $\rho_\mathrm{dr}$ weakly depends on $\Delta$. Therefore the crucial property of the superconducting layer required to support ABE is not a large value of the gap itself, but its critical temperature $T_\mathrm{c}^\mathrm{SC}$, which should be high enough in comparison to the experimental $T$. Second, the dominating contribution to $\rho_\mathrm{dr}$ is provided by the ``condensate'' processes involving transitions between the condensate and noncondensate polaritons induced by their interaction with electrons, shown in Fig. \ref{fig: Diagram for EP}. Similar processes were considered as dominating in the theory of normal drag between excitonic BEC and non-superconducting 2DEG \cite{Boev2019}. The ``noncondensate'' processes typical to the theory of normal Coulomb drag \cite{Rojo1999} provide negligible contribution, as demonstrated in our calculations.

We also show that the important point for reliable theoretical calculations of $\rho_\mathrm{dr}$ is the appropriate screening of the interlayer interaction. Neglect of the screening results in $\rho_\mathrm{dr}$ overestimated by 1-2 orders of magnitude. On the other hand the screening in the widely used Thomas-Fermi approximation \cite{Boev2019,Laussy2010, Laussy2012, Cotlet2016} underestimates $\rho_\mathrm{dr}$ by an order of magnitude. We use the random phase approximation for both electronic and polaritonic layers which provides intermediate, and more realistic, results for $\rho_\mathrm{dr}$. Note that dynamical effects in screened interlayer interaction play an essential role in the theory of normal drag too \cite{Narozhny2016}.

We consider typical experimental conditions for realization of excitonic and electronic layers using GaAs-based semiconductor QWs or two-dimensional TMDC crystals. Spatially indirect dipolar excitons are the most promising for observation of ABE because of their relatively strong interaction with electrons, although achievement of the dipolar polaritonic condensation (BEC of dipolaritons) is still experimentally challenging. At the distance $L$ between excitonic and electronic layers not exceeding 40 nm, $\rho_\mathrm{dr}$ reaches the values of 0.001-0.1 of the total mass density of polaritons $\rho_\mathrm{p}$. The direct excitons interacting with electrons owing to their polarizability provide $\rho_\mathrm{dr}$ which is orders of magnitude lower than for indirect excitons. However, this setup with BEC of direct-exciton polaritons coupled to a superconducting layer could be more feasible from the technical point of view.

The effect of the Rabi splitting $\Omega_\mathrm{R}$ on  $\rho_\mathrm{dr}$ is twofold: higher $\Omega_\mathrm{R}$, from the one hand, helps to stabilize polaritons to ensure the strong-coupling regime, while, from the other hand, reduces $\rho_\mathrm{dr}$ due to increased energy of virtual lower polariton excitations in a broad momentum range contributing to the drag. Our results are presented for zero photon-to-exciton detuning $\delta$, but $\rho_\mathrm{dr}$ moderately increases at $\delta>0$ (because the lower polaritons become more exciton-like and their interaction with electrons is enhanced) and decreases at $\delta<0$. At  $\delta=\Omega_{R}$ the drag density is approximately twice as much as at $\delta=0$. Strong-correlation effects such as softening of the Bogoliubov mode dispersion in the polariton system \cite{Cotlet2016} can also increase $\rho_\mathrm{dr}$ in the vicinity of instability against supersolid transition at high enough polariton density. However, maintaining uniformity of the polariton BEC and its superfluidity can be experimentally challenging in this regime.

For detection of the predicted ABE, we suggest to create a flow of polariton Bose condensate and to detect a supercurrent induced in the electronic layer. At typical polariton velocities $10^8\,\mbox{cm/s}$ and with the values $\rho_\mathrm{dr}\sim(0.001-0.1)\rho_\mathrm{p}$ predicted in our calculations, we expect the superconducting current to be not much smaller that the critical current of conventional superconductors. Thus the predicted effect could be measurable at realistic conditions, if the excitonic layer of polariton (or dipolariton) BEC and superconducting electron gas can be brought to small enough distance. Additionally, we predict similar ABE for a coupled system of electrons and Bose-condensed spatially indirect excitons in the absence of microcavity. Magnitude of the drag density in this case is close to that in electron-polariton system, although observation of ABE should be more difficult due to lower velocities of excitons.

Different mechanisms of electron entrainment by polaritons, such as normal drag \cite{Narozhny2016,Rojo1999}, polaronic effects \cite{Cotlet(2019)}, or photon drag \cite{PhDrag} can, in principle, compete with ABE leading to similar experimental signatures. Nevertheless, the first two effects should affect only a normal component of electronic gas and thus can be separated in measurements of a nondissipative current. The photonic drag could be caused by entrainment of the superconducting Cooper pairs due to transfer of momentum of absorbed microcavity photons, although there are no experimental confirmations of existence of such effect in superconductors.

We analyzed the superfluid drag in a clean system, assuming that both polariton BEC and Cooper-pair condensate in superconducting 2DEG remain approximately uniform even in presence of weak disorder. For the normal Coulomb drag, the role of mesoscopic fluctuations is known to be significant at low temperatures \cite{Narozhny2000,Kim2011,Narozhny2016}, so it would be interesting to study an influence of fluctuations and spatial nonuniformity on the superfluid drag. Fluctuations of the superconducting order parameter can also play important role near the superconducting transition temperature $T_\mathrm{c}^\mathrm{SC}$. The same concerns possible influence of the Berezinskii-Kosterlitz-Thouless transition physics \cite{Caputo2017} and finite-size effects. For example, in magnetic field the boundary conditions for currents can even change the sign of observable Coulomb drag \cite{Titov2013}. Similarly, in our proposed experimental setup in Fig.~\ref{fig:ExpSet} the measured drag current or voltage (or other quantity like magnetic flux) can depend not only on a magnitude of ABE itself, but also on a superconducting circuit incorporating the passive electronic layer.

In our setting, we assumed that superconductivity is induced in the electronic layer by the polaritonic mechanism, which was proposed theoretically in Refs. \cite{Laussy2010, Laussy2012, Cherotchenko2016, Cotlet2016, Petros2018, Sedov2019, Meng2021, Sun2021_1, Sun2021} but not yet observed in the experiments. However, this specific pairing mechanism in not important for the presence of ABE, which is expected also for a preexisting superconducting layer. In the latter case, a possible obstacle for observation of ABE could be detrimental influence of the superconductor on the optical microcavity quality. To minimize it, thin-film or even atomically-thick superconductors, such as $\mathrm{NbSe}_2$ \cite{Xi2015} or FeS \cite{Huang2017}, can be used. Another way is to choose different geometries for a microcavity (for example, an optical fiber \cite{Sedov2019, Petros2018}) or a superconductor (thin wire or grid). Such setup on superfluid drag between BEC excitonic polaritons and a superconductor will be analyzed elsewhere \cite{Aminovtbp}. One more possibility is the $p_x+ip_y$ superconducting pairing suggested in \cite{Cotlet2016,Julku2022} for electron-exciton or electron-polariton system and in \cite{Kinnunen2018} for a similar atomic Bose-Fermi mixture. Our calculations show that the drag density does not change significantly on transition from the s- to $(p_x+ip_y)$-wave pairing, since there is no specific selection of angular harmonics of the gap in the superconductor rectification function (\ref{FelG}); in the case of $(p_x+ip_y)$-wave pairing, $\rho_\mathrm{dr}$ is slightly higher at the same superconducting $T_\mathrm{c}^\mathrm{SC}$ mainly due to higher ratio $\Delta(T=0)/T_\mathrm{c}^\mathrm{SC}$.

The superfluid drag effect considered in this paper couples a quantum coherent system of half-light, half-matter polaritons, and a superconducting system. The first system hosts fast-traveling bosonic quasiparticles controllable by light, and the second one consist of slower fermionic particles, which can be easily controlled by electric and magnetic fields. Therefore coupling the two systems by the superfluid drag can be used for coherent transfer of information in future hybrid quantum devices.

\section*{Acknowledgments}
The work on analytical calculations of the superfluid drag was supported by the Russian Foundation for Basic Research (RFBR) within the Project No. 21–52–12038. The work on numerical calculations was supported by the Program of Basic Research of the Higher School of Economics.

\bibliographystyle{quantum}
\bibliography{Bibtex}

\appendix
\section{Additional diagrams for rectification function of polaritons}\label{Appendix_A}

\begin{figure}[!b]
\centering
\includegraphics[width=0.8\columnwidth]{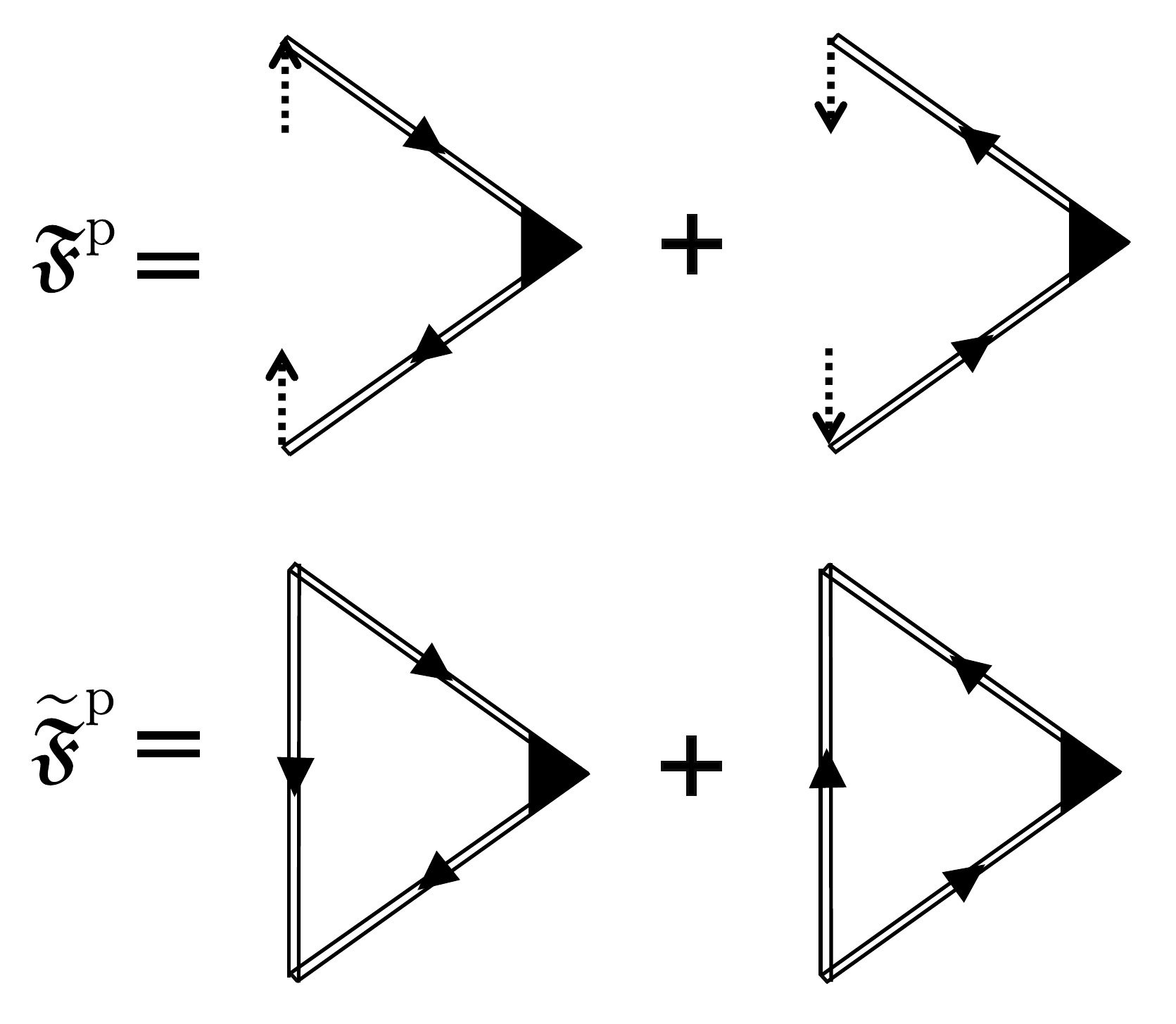}
\caption{Diagrams contributing to $\boldsymbol{\mathfrak{F}}^\mathrm{p}$. The ``condensate'' diagrams in the first line, corresponding to Eq.~(\ref{Fdia}), were taken into account in our calculations of $\rho_{\mathrm{dr}}$, and the ``noncondensate'' diagrams in the second line, corresponding to Eq.~(\ref{Closed}), were neglected.}
\label{fig:DiagramsClosed}
\end{figure}

\begin{figure}[!b]
\centering
\includegraphics[width=\columnwidth]{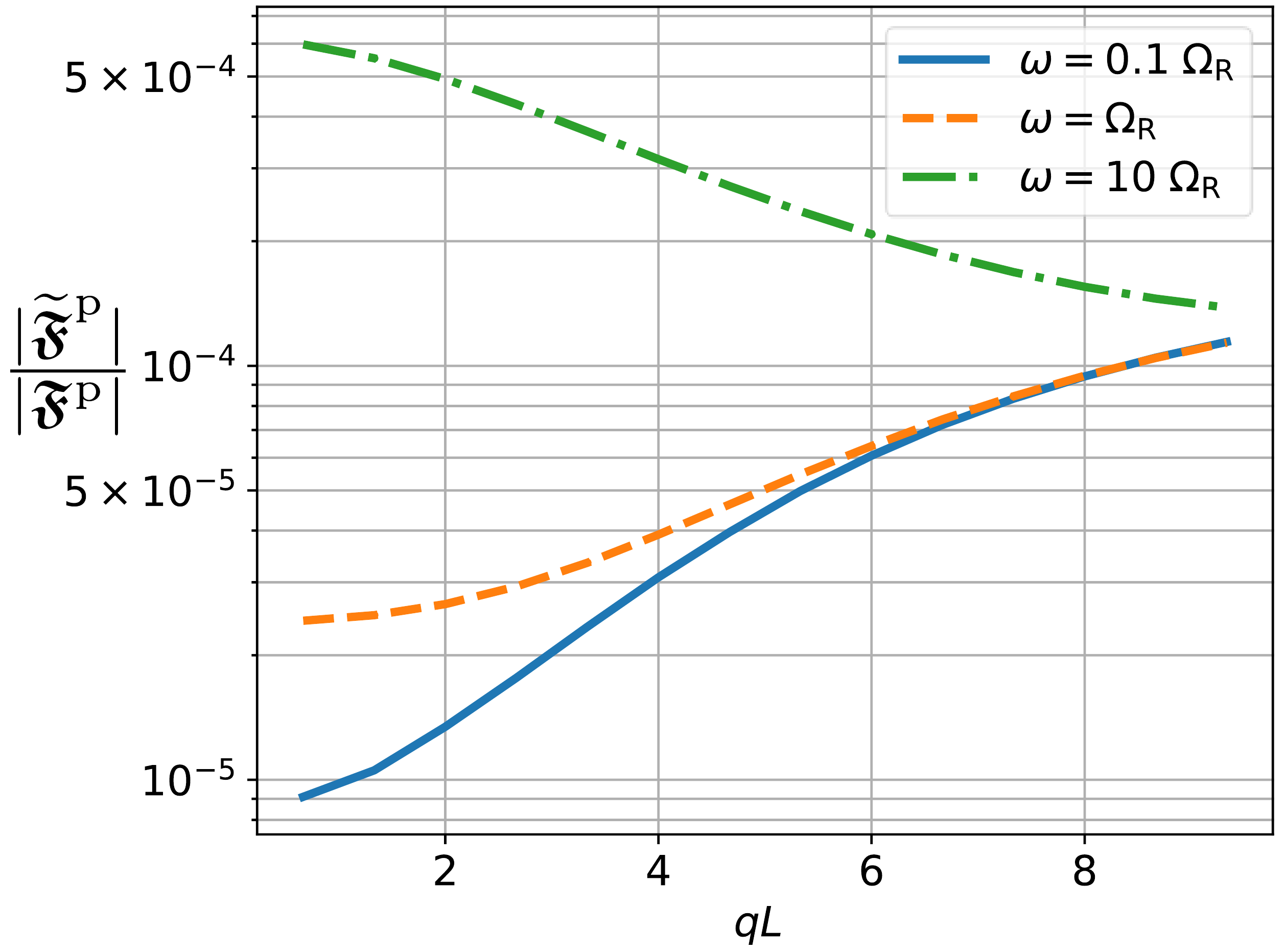}
\caption{Ratio of absolute values of contributions to the rectification function of the polariton system provided by the noncondensate diagrams ($\widetilde{\mathfrak{F}}^{\mathrm{p}}$, second line of Fig. \ref{fig:DiagramsClosed}) and by the condensate diagrams ($\mathfrak{F}^{\mathrm{p}}$, first line of Fig. \ref{fig:DiagramsClosed}). The system parameters are the same as in Fig.~\ref{fig:pol3L}(a) with $\Omega_{\mathrm{R}}=30\,\mbox{meV}$, $L=4\,\mbox{nm}$, both exciton and electron layers are TMDC-based.}
\label{fig:Closed}
\end{figure}

In this Appendix we consider the subleading ``noncondensate'' diagrams for the rectification function of the polariton system, depicted in the lower panel of Fig.~\ref{fig:DiagramsClosed}. Such closed diagrams are usually considered in the theory of Coulomb drag \cite{Narozhny2016} between normal systems, where they provide the leading contribution, although in Bose-condensed systems they may be disregarded \cite{Boev2019}. However, accounting for the noncondensate diagrams can be important in some cases, as argued, e.g., in Ref.~\cite{Meng2021} in the context of polariton-mediated superconductivity. For this reason we analyze a contribution of such diagrams, depicted in Fig.~\ref{fig:DiagramsClosed}, to the rectification function:
\begin{multline}\label{Closed}
\widetilde{\boldsymbol{\mathfrak{F}}}^{\mathrm{p}}(\mathbf{q},i\omega_{m})  = \frac{T}S \sum_{\mathbf{p},i\omega^{\prime}_{n}} \mathbf{p}\,\mathrm{Tr} \left[\hat{G}^{\mathrm{p}} (\mathbf{p},i\omega_{n}^{\prime})\tau_{3}\hat{G}^{\mathrm{p}} (\mathbf{p},i\omega_{n}^{\prime}) \right. \\ \left. \times \hat{G}^{\mathrm{p}} (\mathbf{p+q},i\omega_{n}^{\prime}+i\omega_{m})\right] \\ +  (\mathbf{q},i\omega_{m}\to -\mathbf{q},-i\omega_{m}).
\end{multline}
In the $T=0$ limit, after the frequency summation, this expression takes a form similar to that for the electron rectification function (\ref{Fel}):
\begin{multline}\label{F_noncond}
\widetilde{\boldsymbol{\mathfrak{F}}}^{\mathrm{p}}(\mathbf{q},i\omega_m) \\ =\frac1S \sum_{\mathbf{p}}\frac{\mathbf{p} (E^{\mathrm{p}}_{\mathbf{p}}+E^{\mathrm{p}}_{\mathbf{p+q}})i\omega_m L^{\mathrm{p}}(\mathbf{p},\mathbf{q}) }{\left[(i\omega_m)^2-\left(E^{\mathrm{p}}_{\mathbf{p}}+E^{\mathrm{p}}_{\mathbf{p+q}}\right)^{2}\right]^{2}},
\end{multline}
where $L^{\mathrm{p}}(\mathbf{p},\mathbf{q})= (u^{\mathrm{p}}_{\mathbf{p}}v^{\mathrm{p}}_{\mathbf{p+q}} + u^{\mathrm{p}}_{\mathbf{p+q}}v^{\mathrm{p}}_{\mathbf{p}})^{2}$ is the bosonic coherence factor, and $u_\mathbf{p}^\mathrm{p},v_\mathbf{p}^\mathrm{p} = \pm \sqrt{\frac12\{\pm1+(\tilde\epsilon^\mathrm{p}_{\mathbf{p}} + c_{p} )/E^{\mathrm{p}}_{\mathbf{p}}\}}$ are the coefficients of the polaritonic Bogoliubov transformation.

Fig. \ref{fig:Closed} shows the comparison of the leading-order condensate contribution $\boldsymbol{\mathfrak{F}}^{\mathrm{p}}$ (\ref{Fb}) to the rectification function with the subleading noncondensate one $\widetilde{\boldsymbol{\mathfrak{F}}}^{\mathrm{p}}$ (\ref{F_noncond}). Since the main contribution to the integral in Eq.~(\ref{chidiag}) is provided by $\omega\sim \Omega_{\mathrm{R}}$ and $q\sim L^{-1}$, we calculate the rectification functions in the range $q=(1-10)\times L^{-1}$ and $\omega=(0.1-10)\times\Omega_{\mathrm{R}}$. It can be seen that $\widetilde{\boldsymbol{\mathfrak{F}}}^{\mathrm{p}}$ in this range is 3--4 orders of magnitude smaller than  $\boldsymbol{\mathfrak{F}}^{\mathrm{p}}$. These estimates are made for the TMDC-based setup, and for QW-based system the ratio $|\widetilde{\boldsymbol{\mathfrak{F}}}^{\mathrm{p}}|/|\boldsymbol{\mathfrak{F}}^{\mathrm{p}}|\sim10^{-6}$ is even smaller. Therefore we can assuredly neglect the noncondensate diagrams in our calculations.

\section{Temperature dependence of rectification function of superconductor}\label{Appendix_B}

In this Appendix we provide the full expression for the nonlinear current-density response function of 2D superconductor for $T>0$. It is obtained after the frequency summation in Eq.~(\ref{FelG}):
\begin{widetext}
\begin{multline}\label{26}
\boldsymbol{\mathfrak{F}}^{\mathrm{el}} (\mathbf{q},i\omega_{m}) \\ =g_\mathrm{v}\sum_{\mathbf{p}} \mathbf{p}\left(\left\{ \left[  n_{\textrm{F}}(E_{\mathbf{p+q}}^{\mathrm{el}}) - n_{\textrm{F}}(E_{\mathbf{p}}^{\mathrm{el}}) \right] \left[ (M_{+}^{\mathbf{p,q},i\omega_{m}})^{2} - (M_{-}^{\mathbf{p,q},i\omega_{m}})^{2} \right] + n_{\textrm{F}}^{\prime}(E_{\mathbf{p}}^{\mathrm{el}}) \left[ M_{+}^{\mathbf{p,q},i\omega_{m}} - M_{-}^{\mathbf{p,q},i\omega_{m}} \right] \right\}L^{\mathrm{el}}_{2}(\mathbf{p},\mathbf{q})\right.\\ \left.+\left\{ \left[ 1 - n_{\textrm{F}}(E_{\mathbf{p}}^{\mathrm{el}}) - n_{\textrm{F}}(E_{\mathbf{p+q}}^{\mathrm{el}}) \right] \left[ (P_{+}^{\mathbf{p,q},i\omega_{m}})^{2} - (P_{-}^{\mathbf{p,q},i\omega_{m}})^{2} \right] +  n_{\textrm{F}}^{\prime}(E_{\mathbf{p}}^{\mathrm{el}}) \left[ P_{+}^{\mathbf{p,q},i\omega_{m}} - P_{-}^{\mathbf{p,q},i\omega_{m}} \right] \right\} L^{\mathrm{el}}(\mathbf{p},\mathbf{q})\right)\\
+ (\mathbf{q},i\omega_{m}\to -\mathbf{q},-i\omega_{m}),
\end{multline}
\end{widetext}
where $L^{\mathrm{el}}_{2}(\mathbf{p},\mathbf{q})=(u_{\mathbf{p}}u_{\mathbf{p+q}} - v_{\mathbf{p}}v_{\mathbf{p+q}})^{2}$ is the second fermionic coherence factor; $n_{\textrm{F}}$ and $n_{\textrm{F}}^{\prime}$ are the Fermi-Dirac distribution and its derivative; $P_{\pm}^{\mathbf{p,q},i\omega_{m}}=(E_{\mathbf{p}}^{\mathrm{el}}+E_{\mathbf{p+q}}^{\mathrm{el}} \pm i\omega_{m})^{-1}, M_{\pm}^{\mathbf{p,q},i\omega_{m}}=(E_{\mathbf{p}}^{\mathrm{el}}-E_{\mathbf{p+q}}^{\mathrm{el}} \pm i\omega_{m})^{-1}$.

The energy gap $\Delta$ also depends on the system temperature $T$. This dependence can be approximated by the widely used expression \cite{Gross1986}:
\begin{equation}\label{DeltaT}
\Delta(T)\approx\Delta(0) \tanh \left(  k \sqrt{\frac{T_{\textrm{c}}}{T}-1} \right).
\end{equation}
Here $k=1.74$ and $\Delta(0)=1.76T^{\mathrm{SC}}_{\mathrm{c}}$; these constants correspond only to weak-coupling superconductors with the s-wave pairing, which we are interested in. The formulas (\ref{26}) and (\ref{DeltaT}) are used to calculate the temperature dependence of $\rho_{\mathrm{dr}}(T)$ shown in Fig.~\ref{fig:polT}.

\end{document}